\documentclass{sig-alternate}
\usepackage{float}
\usepackage{url}
\usepackage[font=small]{caption}
 \usepackage{helvet}
\usepackage[lofdepth,lotdepth]{subfig}

\newcommand{\ignore}[1]{}

\usepackage{etoolbox}
\makeatletter
\patchcmd{\maketitle}{\@copyrightspace}{}{}{}
\makeatother

\begin{document}

\title{Differential Data Analysis for Recommender Systems}
\subtitle{Extended version of RecSys 2013 paper}

\numberofauthors{4}
\author{
\alignauthor Richard Chow\titlenote{Work performed while at Samsung Electronics R\&D.}\\
   \affaddr{Intel Corporation}\\
   \email{richard.chow@intel.com}
\alignauthor Hongxia Jin\\
   \affaddr{Samsung Electronics R\&D}\\
   \email{hongxia.jin@samsung.com}
\alignauthor Bart Knijnenburg\raisebox{9pt}{$\ast$}\\
   \affaddr{UC Irvine}\\
   \email{bart.k@uci.edu}
\and
\alignauthor Gokay Saldamli\\
   \affaddr{Samsung Electronics R\&D}\\
   \email{gokay.s@samsung.com}
}


\maketitle

\begin{abstract}
We present techniques to characterize which data is important to a recommender system and which is not. Important data is data that contributes most to the accuracy of the recommendation algorithm, while less important data contributes less to the accuracy or even decreases it.  Characterizing the importance of data has two potential direct benefits: (1) increased privacy and (2) reduced data management costs, including storage. For privacy, we enable increased recommendation accuracy for comparable privacy levels using existing data obfuscation techniques. For storage, our results indicate that we can achieve large reductions in recommendation data and yet maintain recommendation accuracy. 
 
Our main technique is called differential data analysis. The name is inspired by other sorts of differential analysis, such as differential power analysis and differential cryptanalysis, where insight comes through analysis of slightly differing inputs. In differential data analysis we chunk the data and compare results in the presence or absence of each chunk. We present results applying differential data analysis to two datasets and three different kinds of attributes. The first attribute is called user hardship. This is a novel attribute, particularly relevant to location datasets, that indicates how burdensome a data point was to achieve. The second and third attributes are more standard: timestamp and user rating. For user rating, we confirm previous work concerning the increased importance to the recommender of data corresponding to high and low user ratings.
\end{abstract}
\ignore{
}

\section{Introduction}
Cloud services that use recommender systems have become increasingly common
and most of these systems exist by virtue of ``big data'': mind-boggling amounts of data are stored and used for recommendation and personalization purposes. One example is
location recommender systems (e.g., recommending nearby points-of-interest), a large and growing area because of the connection with mobile devices.
However, with the increased attention to privacy of user data, many users are uncomfortable with their data being held in the cloud where it could be sold, stolen, or misused~\cite{awad}. User location is particularly sensitive~\cite{PEW2}.

Even the service providers only reluctantly store this user data, despite its tremendous business value. One reason is that the data is a liability: it must be protected and sub-poenas must be responded to. Inadvertent data leaks, hacks, or release of data to government agencies can be a public relations problem. Another reason is the volume - ``big'' data is increasing 60\% or more a year~\cite{BigDataStorage}. Storage and management costs are encouraging advances in technologies such as data deduplication, currently a multi-billion dollar market.

This begs the question: Is all this data really necessary for making good recommendations? If not, what part of the data can be safely discarded while maintaining recommendation quality? This question is one motivation for this paper, but another motivation comes from the observation that reducing the amount of data on the server does not address user privacy concerns. A natural and well-studied approach to enhance user privacy is to obfuscate before upload.
How do we minimize the impact to the recommender of this obfuscation?

Our approach starts with techniques to rank data according to its usefulness to the recommender system. In essence, we identify attribute values which are associated with the usefulness of the data to the recommender. We test the association through what we call {\it differential data analysis}: we select data according to attribute values and observe the effect of the presence and absence of this data on recommendation accuracy. 

Ranking of data has applications on both the client and server side. On the server side, we present techniques for a recommender server to filter data by rank. These techniques reduce the amount of data and yet maintain recommendation accuracy. Data can be filtered as it arrives, or there may be initial and periodic training periods, in which some or all data is kept unfiltered in order to learn or update filtering parameters. After filtering, the remaining data is stored by the recommender for use. Data that is only marginally useful to the recommender can be deleted by the system (or even not uploaded by the client). The main advantage of our technique is that it can significantly reduce the amount of data needed by the recommender and yet maintain nearly the same recommendation accuracy. 

On the client side, one key application area for our techniques is user privacy. A system could use our techniques to enable clients, {\em before uploading,} to rank the usefulness of data to the server and then use standard privacy-enhancing techniques such as obfuscation by adding fake data and/or filtering data. These obfuscation techniques are well-studied in the literature. Manual obfuscation is also popular and intuitive among users. Even in 2000~\cite{PEW2000}, 24\% of Internet users provided fake information to a Web site. User studies in~\cite{Benisch,Toch,Locaccino} showed that user filters, for instance turning off upload of location traces from 10 pm -- 6 am or outside of certain regions, ``play an important role in capturing people's privacy preferences.'' We show how these kinds of techniques might affect recommendation accuracy, and in general we show how to optimize the filtering of actual data and/or adding of fake data so that recommendation accuracy is affected least.

\section{Differential Data Analysis}
Our main experimental technique for characterizing the data with respect to its effect on recommendation accuracy is {\it differential data analysis}. At a high level, we divide the data into chunks and compare results in the presence or absence of each chunk. More specifically, suppose a data attribute ranks each user's data in some way, for instance by time or user rating. One can then divide the data into chunks based on this ranking. For convenience, we often divide into 10 chunks, or {\it deciles}. We can then examine the relative effect of a particular decile on recommendation accuracy as follows. We form a training and test set as usual. We rank each user's training data by the data attribute under examination. We remove the first decile of each user's data from the training set and calculate the recommendation accuracy (using a fixed algorithm). We continue by removing the second decile of each user's data, etc., and we end up with 10 readings for the accuracy, which can then be compared. A relatively high accuracy reading implies the corresponding decile is less important to the recommender; a low accuracy reading implies the corresponding decile is more important.

Note that with this method the results may depend on the particular recommendation algorithm employed. In our experiments we found that results did not change when we used algorithms from the same algorithm family, for example, matrix factorization algorithms in Section~\ref{sec:ML}. We also do not expect much change when the algorithms give nearly the same recommendations, for instance, when both are close to ``optimal.''

Note also that there are other ways to do the differential data comparison. Rather than removing data deciles from all users simultaneously, an alternative approach would consider each user in isolation. Deciles would be removed from one user at a time, the effect on recommendation accuracy would be measured for that user, and accuracy measurements would be aggregated over all users at the end. For the experiments described in this paper, we used the former approach since it is less computationally intensive.

\section{Data Attributes}
We study three data attributes in this work. The first attribute we examine is a ``user hardship'' for each data point, the effort it takes the user to attain the data point. The idea is that data points which are more or less to achieve a data point may be more or less indicative of a user's preferences.
This attribute is most intuitively associated with location datasets. To compute the user hardship on our location dataset, a user's location traces can be clustered according to surface-of-the-earth distance, and points are ranked according to their distance to the set of cluster centroids. We call this the KMeans user hardship measure. Points furthest away from the cluster centroids have a higher user hardship score, as these locations are further from a user's usual haunts and require more work to get to. Another way to compute the user hardship measure is to measure the distance to any other point (rather than the cluster centroid). We call this the Density user hardship measure. We present experimental results in Section~\ref{sec:location} on user hardship for a location dataset.

Besides location, a recent study~\cite{Locaccino} found that the time of visit to a location was an indication of privacy sensitivity.
Hence, the second attribute we investigate is the timestamp of the data, for instance the time of a location checkin or the time a rating is given. We investigate whether certain time periods of the day might correspond to more or less important data. In Sections~\ref{sec:location} and~\ref{sec:ML} we present our results on filtering by timestamps for a location dataset and a rating dataset.

Finally, our third attribute is one on which there has been previous work: the actual user rating, for instance the number of stars given by a user for a product. Previous work has recognized that data with high and low ratings are more important to the user~\cite{shardanand} and to the recommender~\cite{Berkovsky}.

\section{Experimental Results}
We experimented with our techniques using two actual datasets, a location dataset and a movie rating dataset. To measure the effect of various filters or fake data on recommendation accuracy, we adopt an application-specific definition of recommendation accuracy, as in~\cite{agrawal}. For each dataset, we use standard recommender algorithms and measure the effect of the filters on the output of these algorithms. For example, for movie ratings prediction, we measure the RMSE with and without the filter for the same algorithm.

\subsection{Location Dataset}
\label{sec:location}

We modeled a location recommender service using an actual database of Gowalla check-ins in the United States collected from June to October 2010\footnote{We thank Betim Berjani and Thorsten Strufe for sharing this dataset.}.
The dataset consisted of 104,875 users and 4,744,089 total locations visited over all users. We eliminated infrequent and inactive users and confined ourselves to check-ins in from several cities with high Gowalla activity. We studied mainly Austin, Texas, which was Gowalla headquarters and had the highest amount of activity. We also studied three other cities: Los Angeles, New York, and Dallas (see Table~\ref{tab:gowalla}).

We considered a simple scheme of each user giving a binary positive rating to each location visited. Locations not visited were considered unrated. We ignored the number of visits. In fact, in our dataset, over 80\% of the ratings were the result of a single visit by a user, i.e. checking in multiple times to the same location was less common than checkin in once. To give users a realistic chance of checking into any location, we confined ourselves to one city at a time. For example, in Austin, we were left with 4,871 users and 245,153 ratings. The number of locations in Austin was 9,577.

\begin{table}
\small
\centering
\begin{tabular}{|l|c|c|c|}
\hline
City     & \# Users & \# Ratings & \# Locations \\
       \hline\hline
Austin   & 4871 & 245153    & 9577\\
\hline
Dallas   & 2991 & 103569    & 5473\\
\hline
Los Angeles & 2957 & 61891  & 10151\\
\hline
New York & 3213 & 106125    & 16861\\
\hline
\end{tabular}
\caption{Gowalla statistics for four cities studied.}
\label{tab:gowalla}
\end{table}

\begin{figure*}[ht]
\begin{minipage}{0.24\textwidth}
\centering
\includegraphics[width=\textwidth]{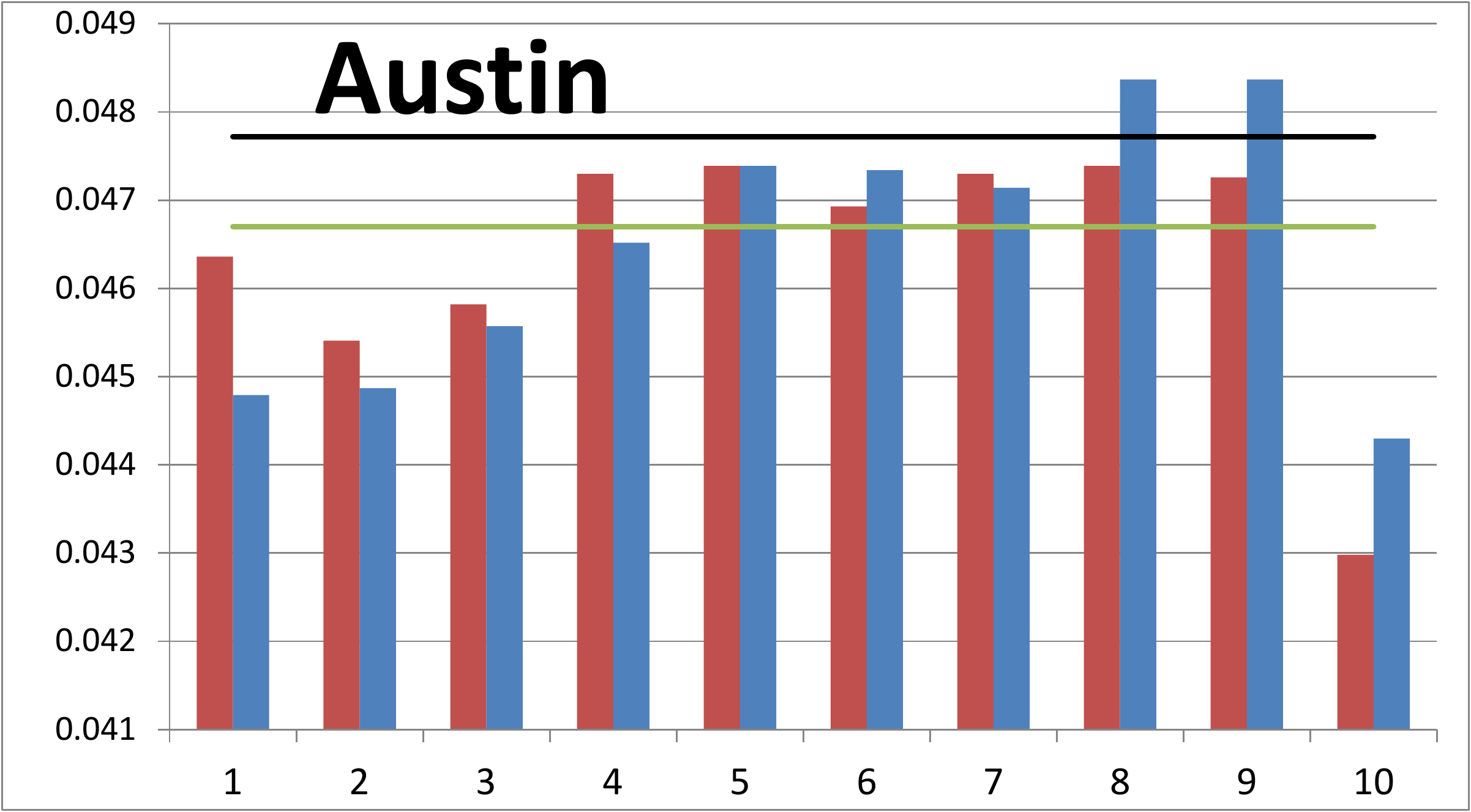}
\end{minipage}
\begin{minipage}{0.24\textwidth}
\centering
\includegraphics[width=\textwidth]{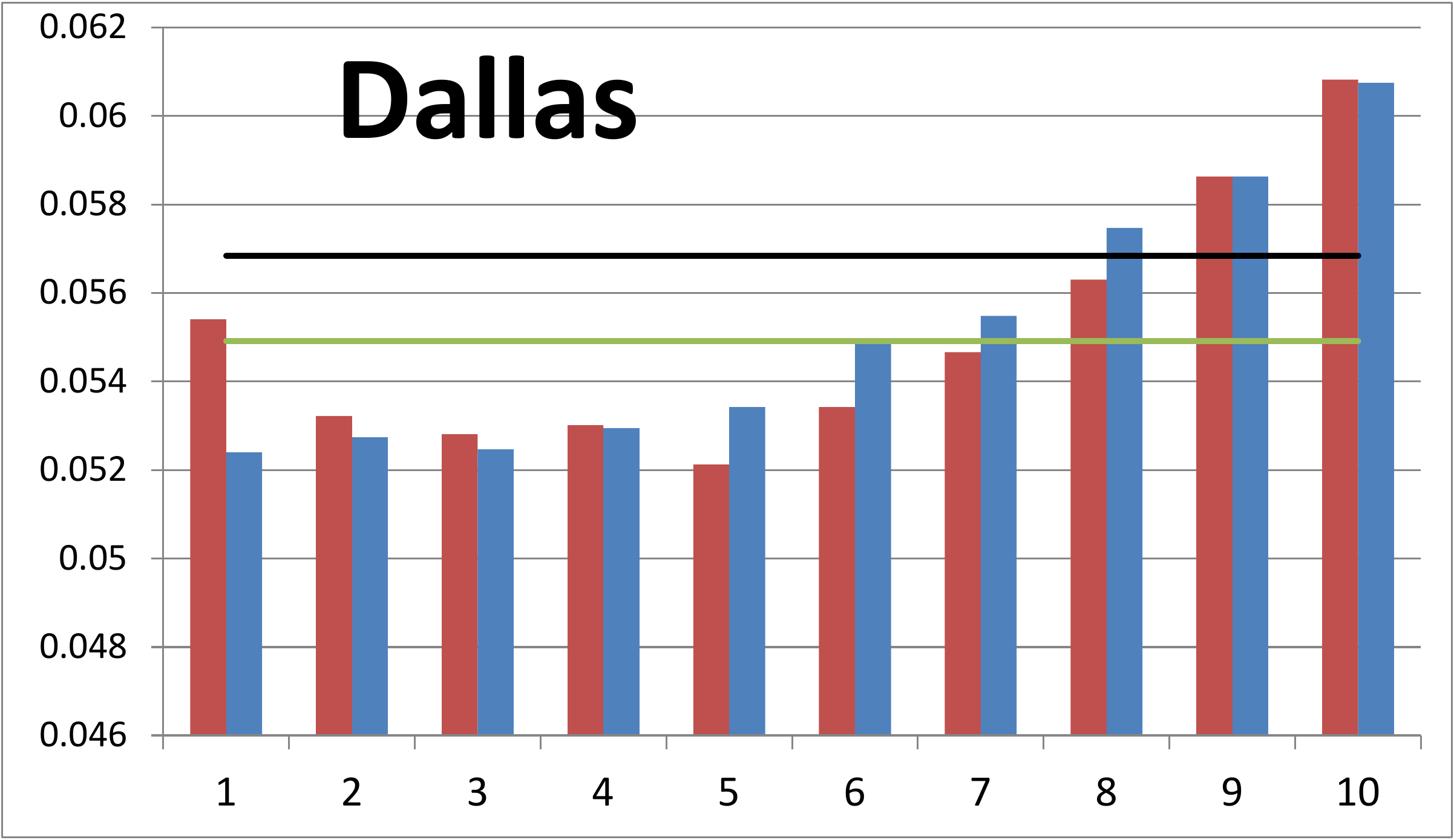}
\end{minipage}
\ignore{
\qquad\qquad
\begin{minipage}{0.30\textwidth}
\centering
\includegraphics[width=\textwidth]{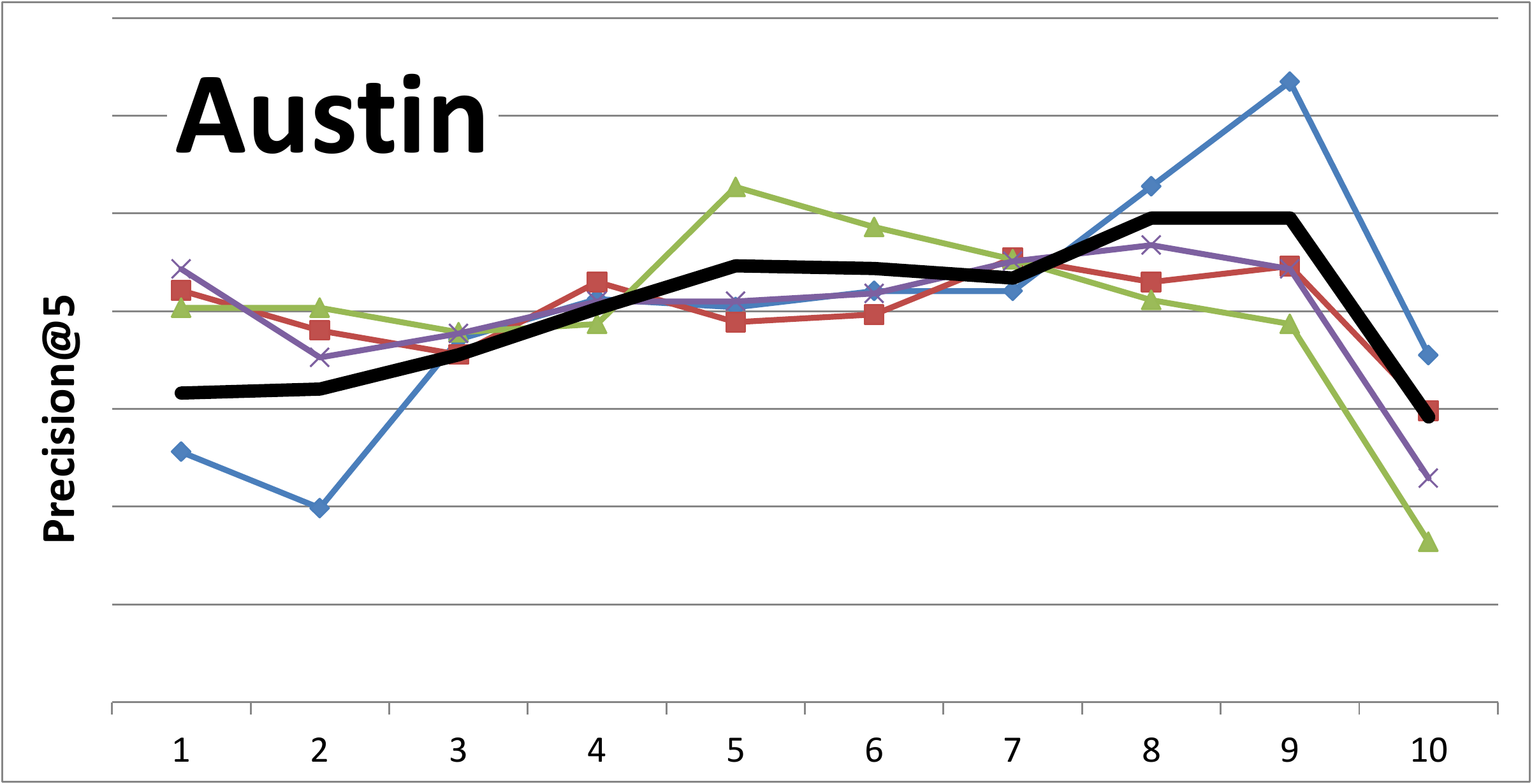}
\label{fig:userstability}
\end{minipage}
}
\begin{minipage}{0.24\textwidth}
\centering
\includegraphics[width=\textwidth]{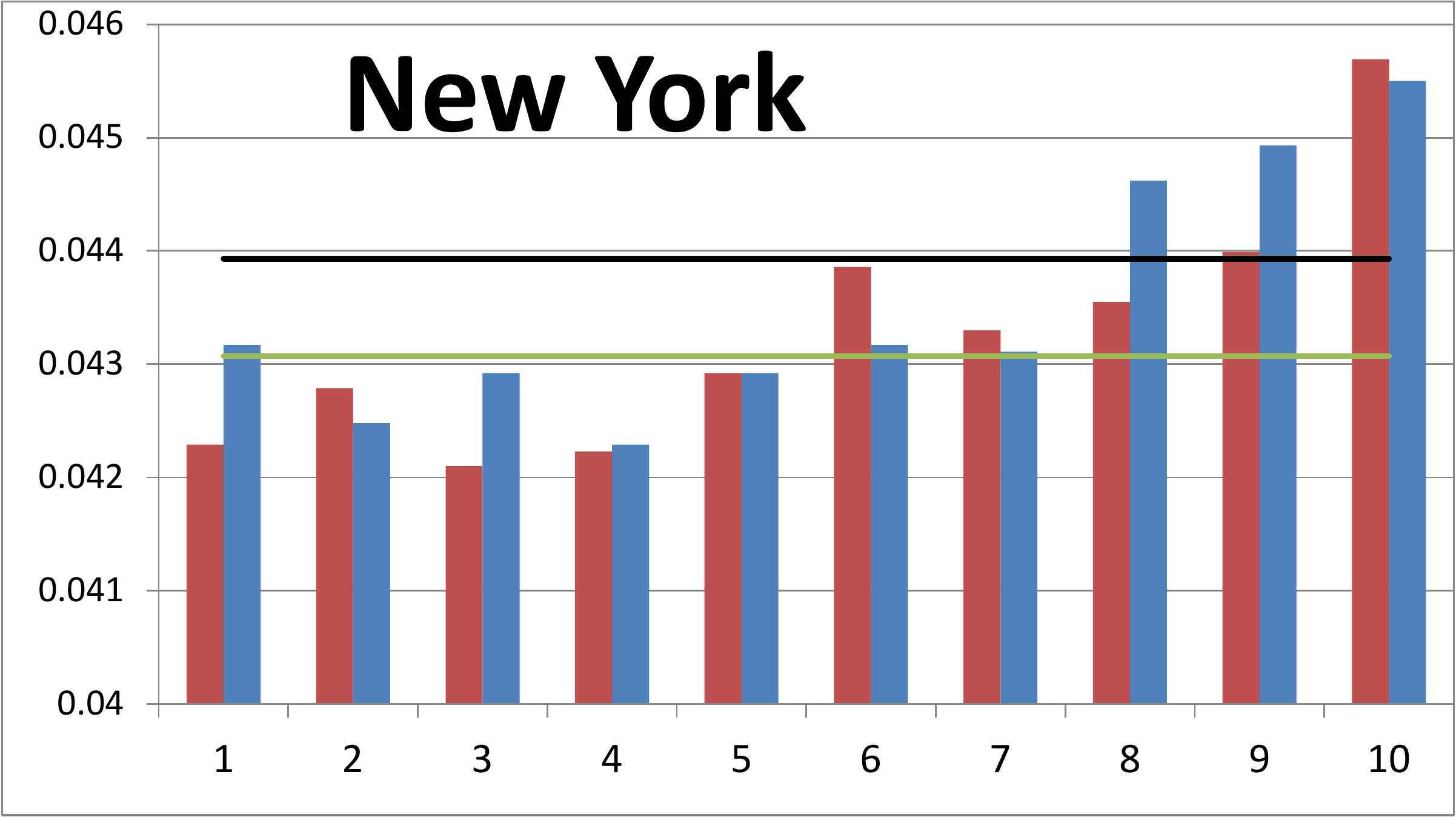}
\end{minipage}
\begin{minipage}{0.24\textwidth}
\centering
\includegraphics[width=\textwidth]{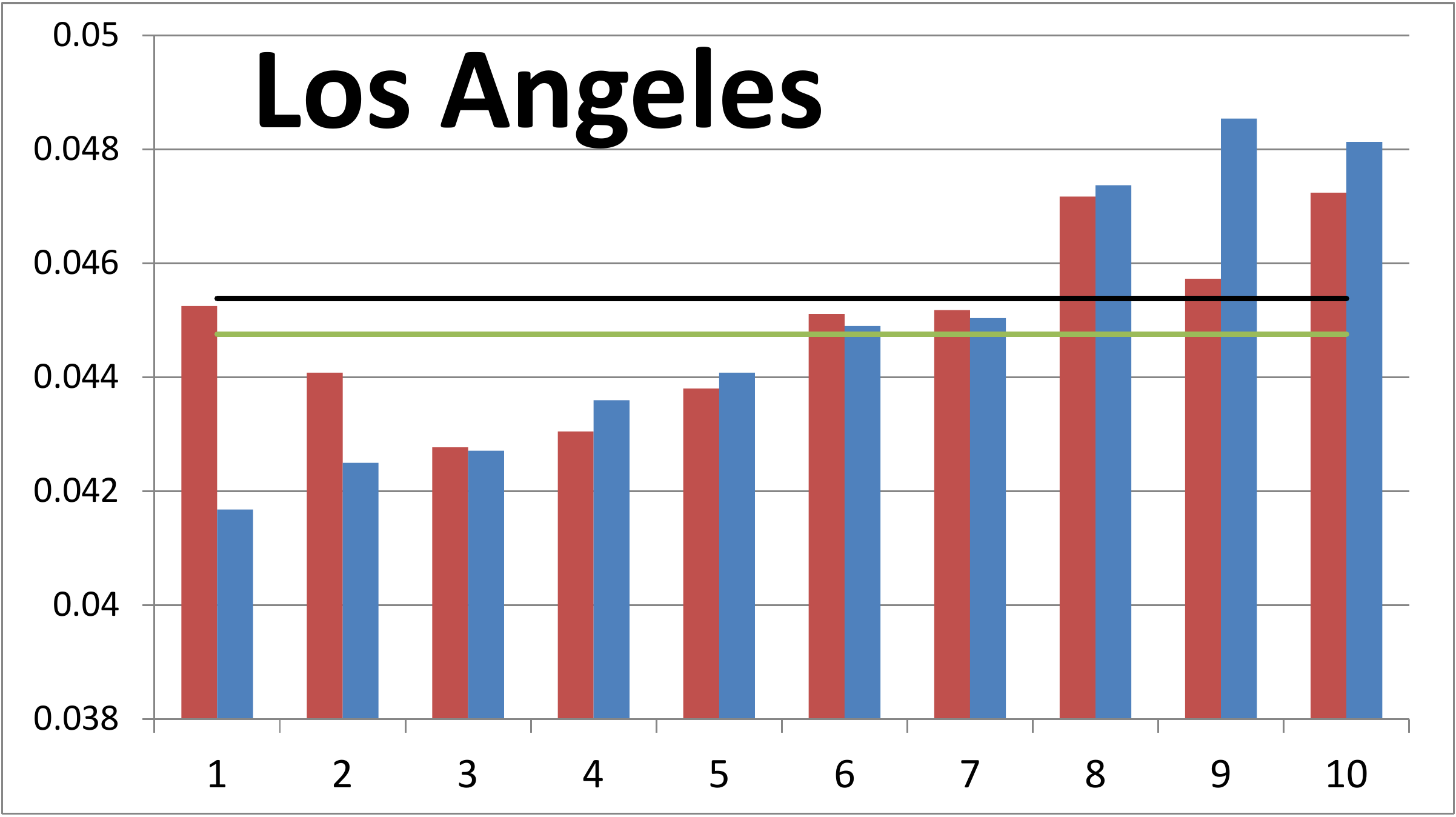}
\end{minipage}
\ignore{
\qquad\qquad
\begin{minipage}{0.30\textwidth}
\centering
\includegraphics[width=\textwidth]{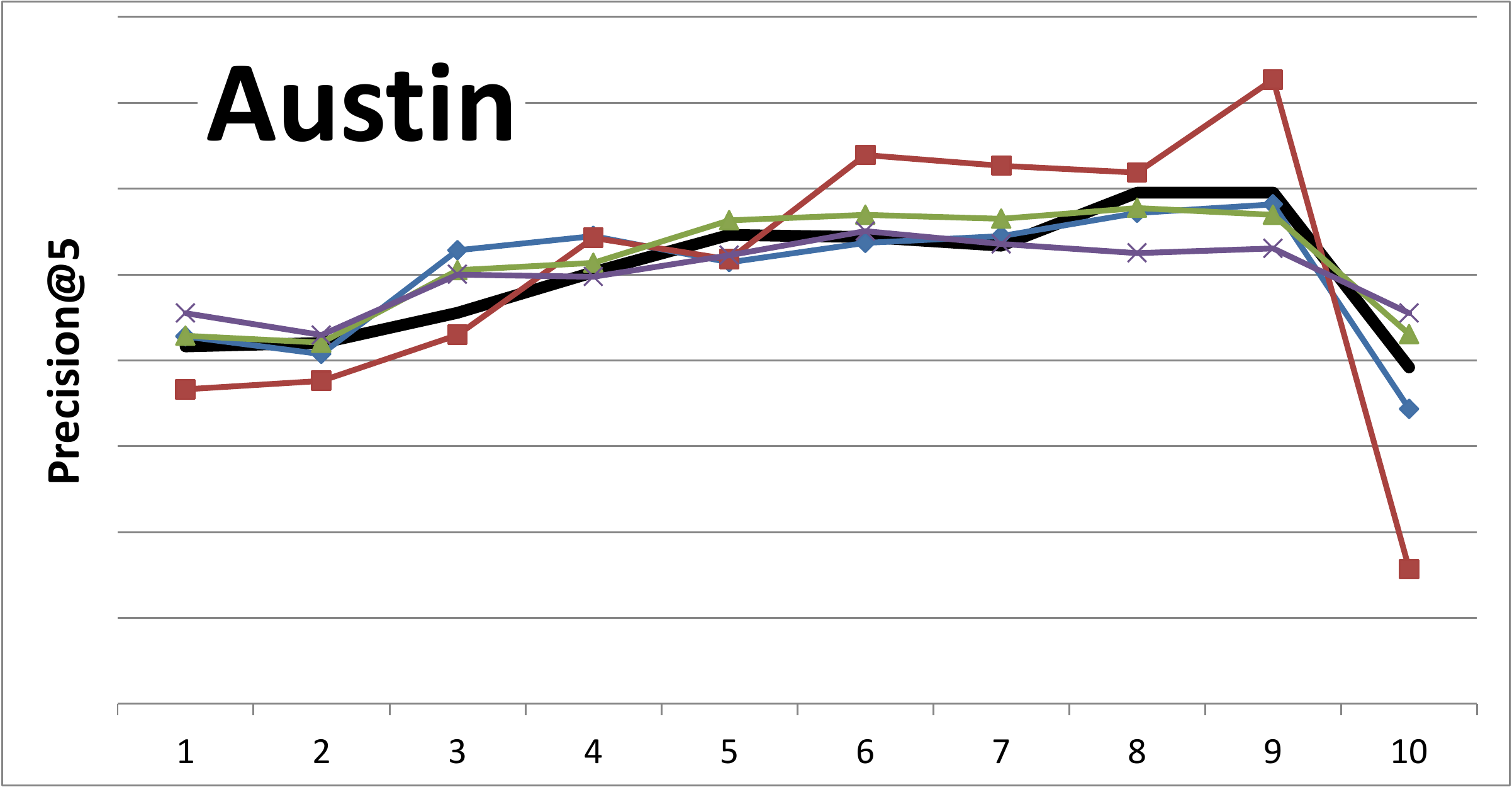}
\label{fig:datastability}
\end{minipage}
}

\begin{minipage}{\textwidth}
\centering
\includegraphics[width=0.25\textwidth]{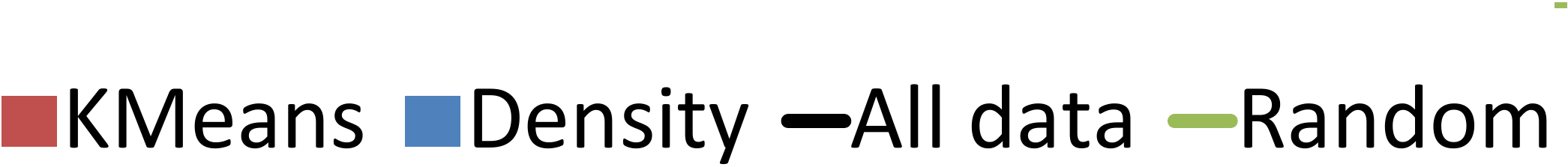}
\end{minipage}

\caption{Recommendation accuracy when each decile (measured by user hardship) is omitted from the training set. Two measures of user hardship are shown, using distance to a user's KMeans centroids and using density at the point. Also shown are lines corresponding to accuracy with the entire training set and with random removal of 10\% of the data. The figures show that the most useful data for recommendation are generally the first few deciles, i.e. the locations not far from where a user usually spends time.}
\label{fig:distance}
\end{figure*}

We randomly chose 20\% of each user's ratings as a held-out test set (e.g., 49,008 test ratings for Austin). We trained a Top-N recommender based on the remaining ratings. The recommender's task is to predict the likelihood of the remaining (user, location)-pairs. For Austin, this is about 46.5 million possible ratings of which 49,008 are test ratings (visited locations).

We used the following standard algorithm for computing recommendations, following~\cite{Ye}. We did not try to optimize the algorithm; our goal was not to achieve the best possible accuracy, but to choose a representative algorithm and see the effect on accuracy of various data attribute values. For each user, we form a binary vector $\mathbf{u}$ where $\mathbf{u}[i] = 1$ if the user has visited location $i$ and $\mathbf{u}[i] = 0$ otherwise. From~\cite{Ye}, we calculate the cosine similarity between any two users $u$ and $v$ by
$$w_{u,v} = \frac{\mathbf{u}\cdot\mathbf{v}}{\|\mathbf{u}\| \|\mathbf{v}\|}.$$

The normalized similarity measure $c_{\mathbf{u},i}$ of $u$ to location $i$ is given by the fraction of users who checked into location $i$ weighted by similarity:
$$c_{\mathbf{u},i} = \frac{\sum_{\mathbf{v} \in L_i}{w_{\mathbf{u},\mathbf{v}}}}{\sum_{\mathbf{v}}{w_{\mathbf{u},\mathbf{v}}}},$$
where $L_i$ are the users who have rated location $i$.

We took the user's top $N$ predicted locations, as given by our similarity measure, and calculated the number of hits, i.e. the number of locations in the test set in this set of $N$ locations. We measured performance using precision and recall. Precision is the fraction of recommendations that are hits; precision@5 means the precision with 5 recommendations.
The recall is the number of hits out of the number of possible hits (i.e., the size of the test set for the user). We use the macro-averaged recall, the average over all users of each user's recall.
As explained in~\cite{Ye}, low precision and recall numbers are expected with such a Top-N recommender. For Austin, our recommendations are an order of magnitude better than random recommendations: five random recommendations would have a precision of approximately $5*(49008/46500000) \approx 0.005$.

\medskip
{\noindent \bf \em User Hardship.}
We considered two measures of user hardship. In the first, we used KMeans (with two centroids, modeling home and work/school) to cluster each user's training points. We then ranked each user's training points according to their minimum distance to the centroids. In the second, for every point, we calculated the minimum distance to any of the user's other training points. We call the two user hardship measures ``KMeans'' and ``Density''. For each measure, we used differential data analysis to rank a user's training points according to the measure: we divided into deciles, and omitted each decile in turn to see the effect on accuracy.

Some of our results are displayed in Figure~\ref{fig:distance}. For clarity and lack of space, we take precision@5 as our proxy for accuracy. We found that recall and precision behaved very similarly in our experiments. Also similar were precision and recall with $N=10$ and $N=20$ recommendations.

Observe that user hardship segregates the data well with respect to effect on recommendation accuracy. With the Austin data, we did 20 trials of randomly removing 10\% of the training data and computed a mean and standard deviation. The ten sample values of our precision@5 statistic ranged from $-3.76$ to $2.60$ standard deviations away from the mean of the random removals, and only 3 of the 10 sam-
ple values were within one standard deviation. In fact, by removing some deciles the accuracy actually became significantly {\it better} than with all the data, implying these deciles have the effect of noise.

The general trend in the four cities we tried was that the lower hardship deciles (i.e., deciles 2, 3, and 4) were most important for accuracy and higher hardship deciles (i.e., deciles 8, 9, 10) were least important and even could be considered noise. One intuitive explanation is that low- to mid-user hardship is the optimum zone for discovering user preferences. A user would not usually endure high user hardship without other reasons besides just his preferences. Austin is a notable exception for the last decile.
The Density user hardship measure generally spreads data points better than KMeans, perhaps owing to the fact that location traces are not defined by only two centroids for many users. We expect the two measures to become closer as the number of KMeans centroids increases.

\medskip
{\noindent \bf \em Timestamp.}
The Gowalla dataset consists of timestamped checkins, so we also tested whether the timestamp attribute could be used to predict the importance of data for the recommender. These timestamps are in local time of the user. We used our technique of differential data analysis and divided up the day into time intervals so that removal of checkins for each interval corresponds to removing about 10-15\% of the data from the training set. We chose this granularity of time interval so one can easily compare with removing 10\% of the data randomly or with removing a decile of data using some other attribute. Note that in our experiments we used a binary measure of the user's preference, so removal of a checkin from one time interval does not affect the training set in the case that the user has checked into the same location in a different time interval. For simplicity, we did not consider the day of the week.

\begin{figure*}[ht]
\begin{minipage}{0.24\textwidth}
\centering
\includegraphics[width=\textwidth]{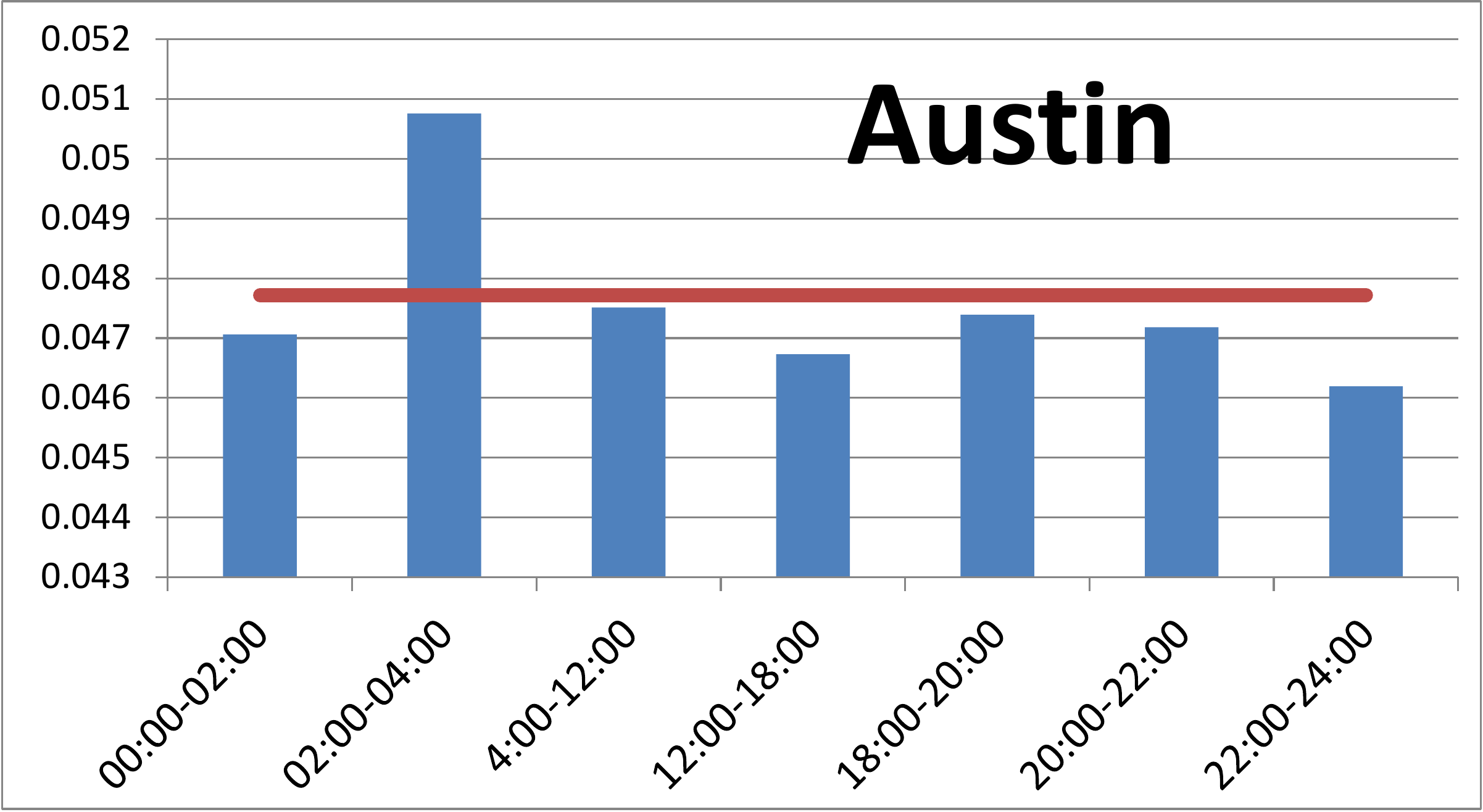}
\end{minipage}
\begin{minipage}{0.24\textwidth}
\centering
\includegraphics[width=\textwidth]{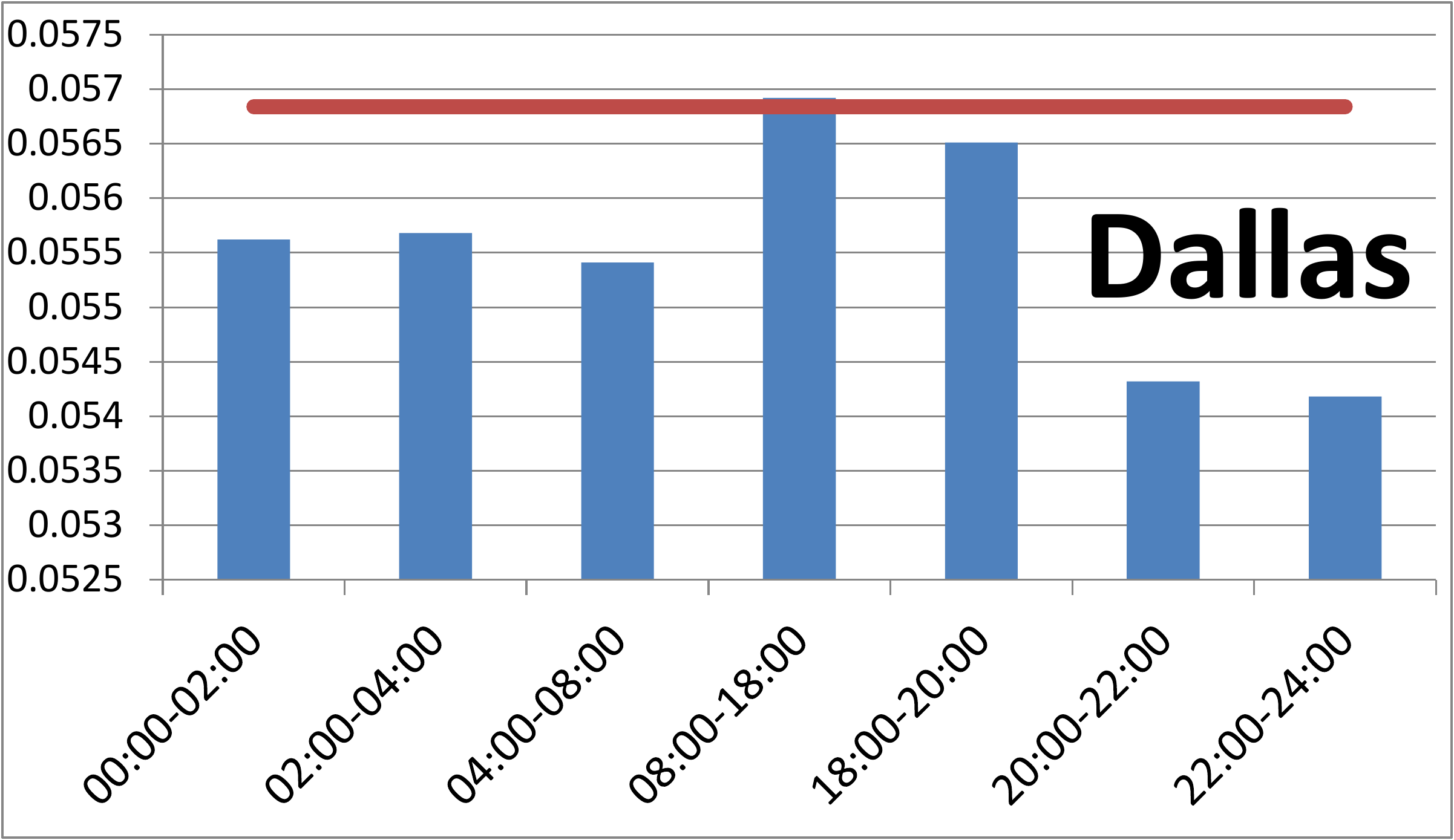}
\end{minipage}
\begin{minipage}{0.24\textwidth}
\centering
\includegraphics[width=\textwidth]{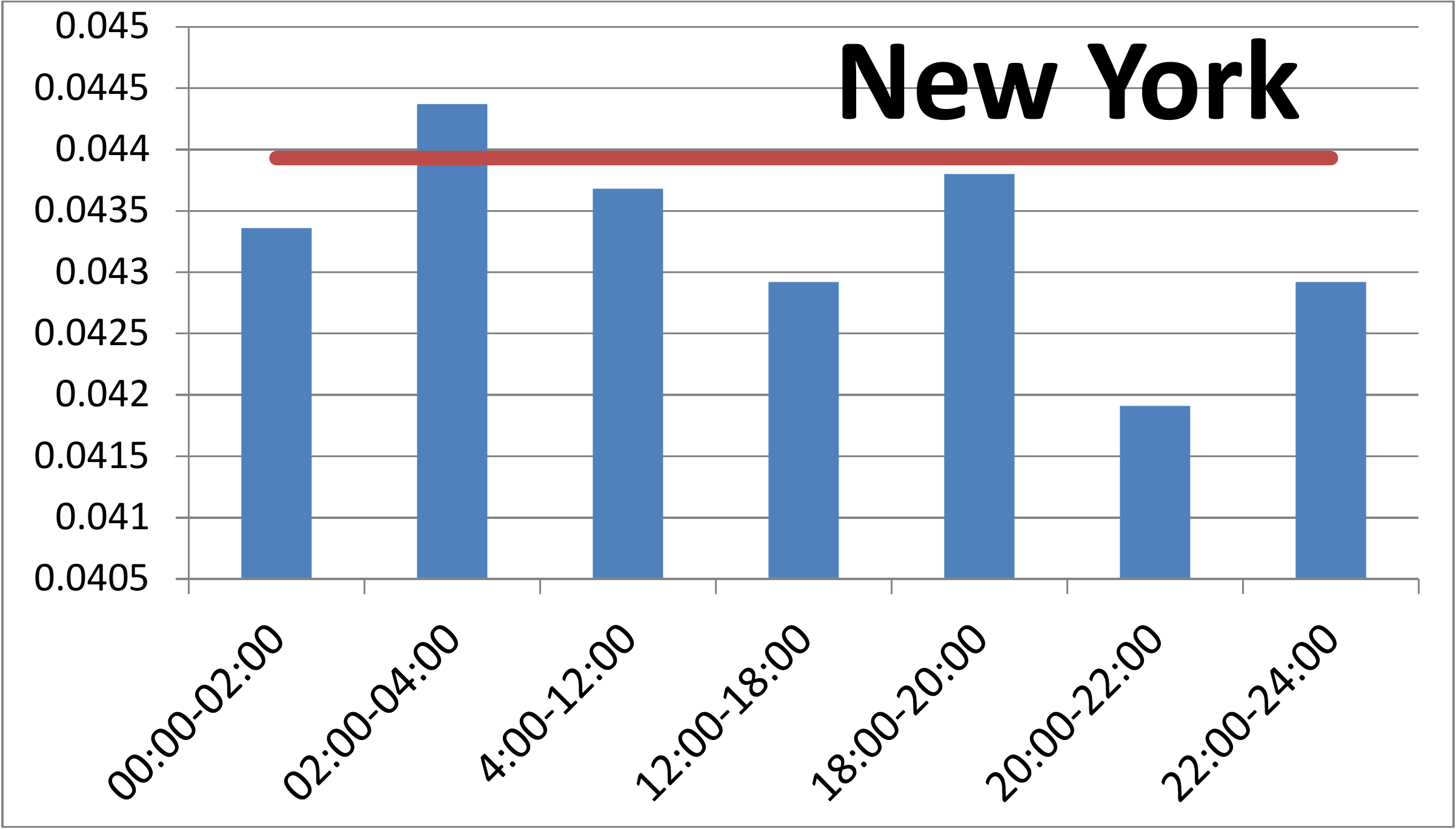}
\end{minipage}
\begin{minipage}{0.24\textwidth}
\centering
\includegraphics[width=\textwidth]{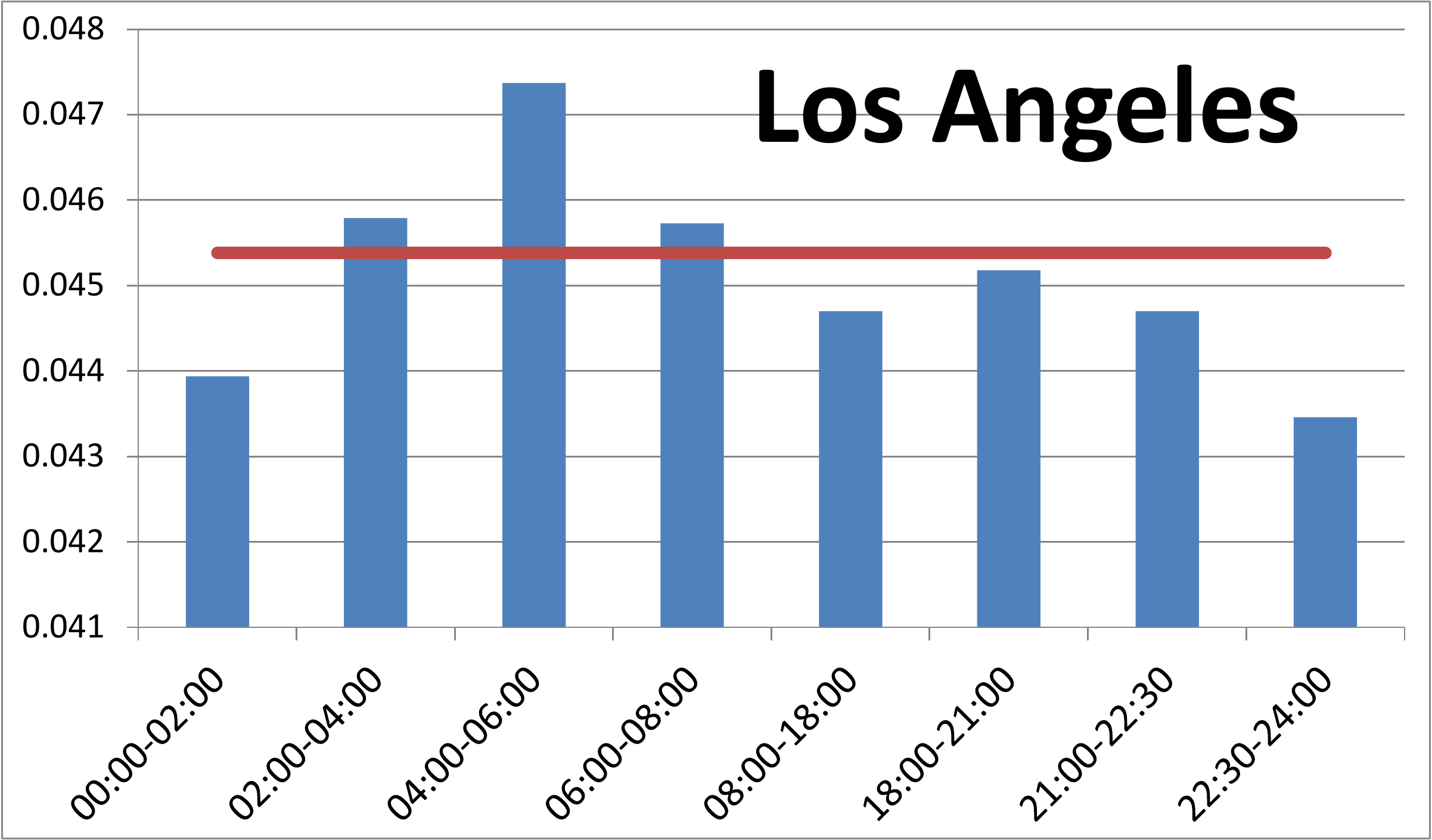}
\end{minipage}

\begin{minipage}{\textwidth}
\centering
\includegraphics[width=0.25\textwidth]{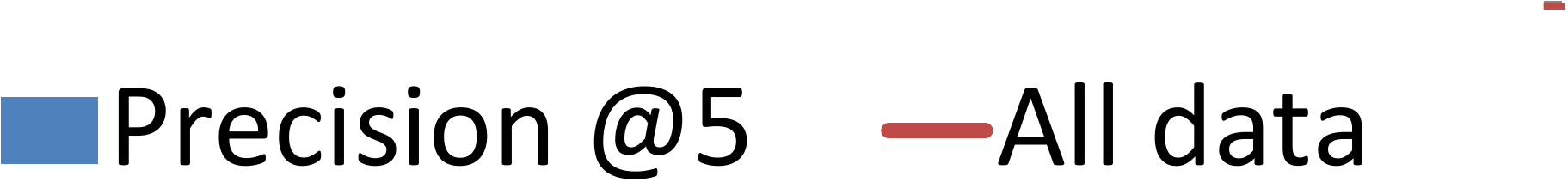}
\end{minipage}
\caption{Recommendation accuracy when distinct time intervals are omitted from the Gowalla training set. Each interval corresponds to about 10-15\% of all data.}
\label{fig:GowallaTime}
\end{figure*}

Our results are shown in Figure~\ref{fig:GowallaTime}. Overall, timestamp seems less predictive of importance than user hardship, although there are some intriguing findings. The most useful data seems to be around 8 p.m. -- midnight, and it appears that post-2 a.m. data is least useful. In three out of the four cities, the data from 2 a.m. -- 4 a.m. seems to even confuse the recommender and decreases accuracy. There are also notable differences in the cities, perhaps related to culture and geography (at least for Gowalla users). For instance, the more important and less important data for Los Angeles seems to come a few hours later than for the other cities.

\begin{figure*}[ht]
\centering
\subfloat[][Different user sets]
{\includegraphics[width=0.30\textwidth]{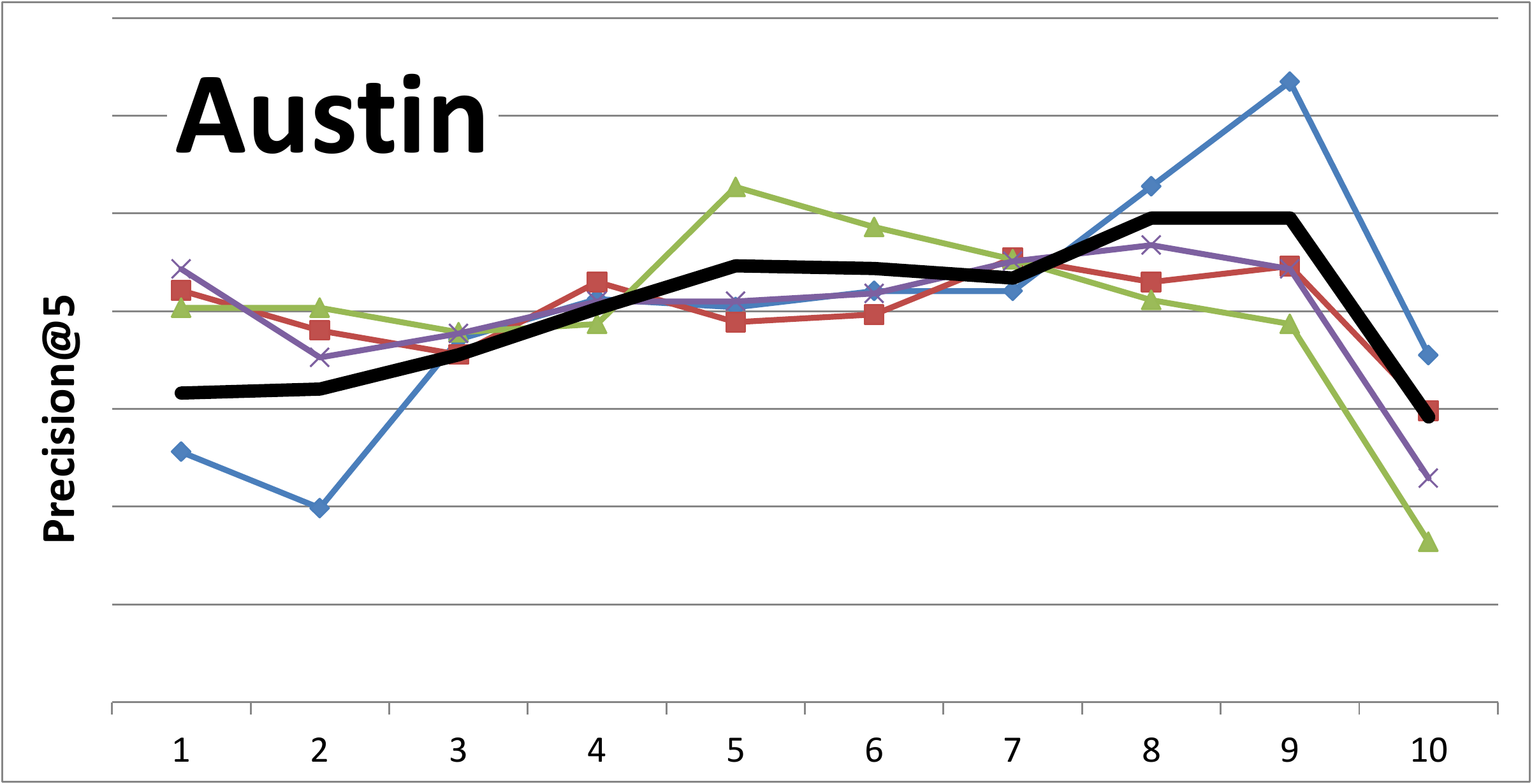}
\label{fig:userstability}}
\qquad
\subfloat[][Different test and training sets]
{
\includegraphics[width=0.30\textwidth]{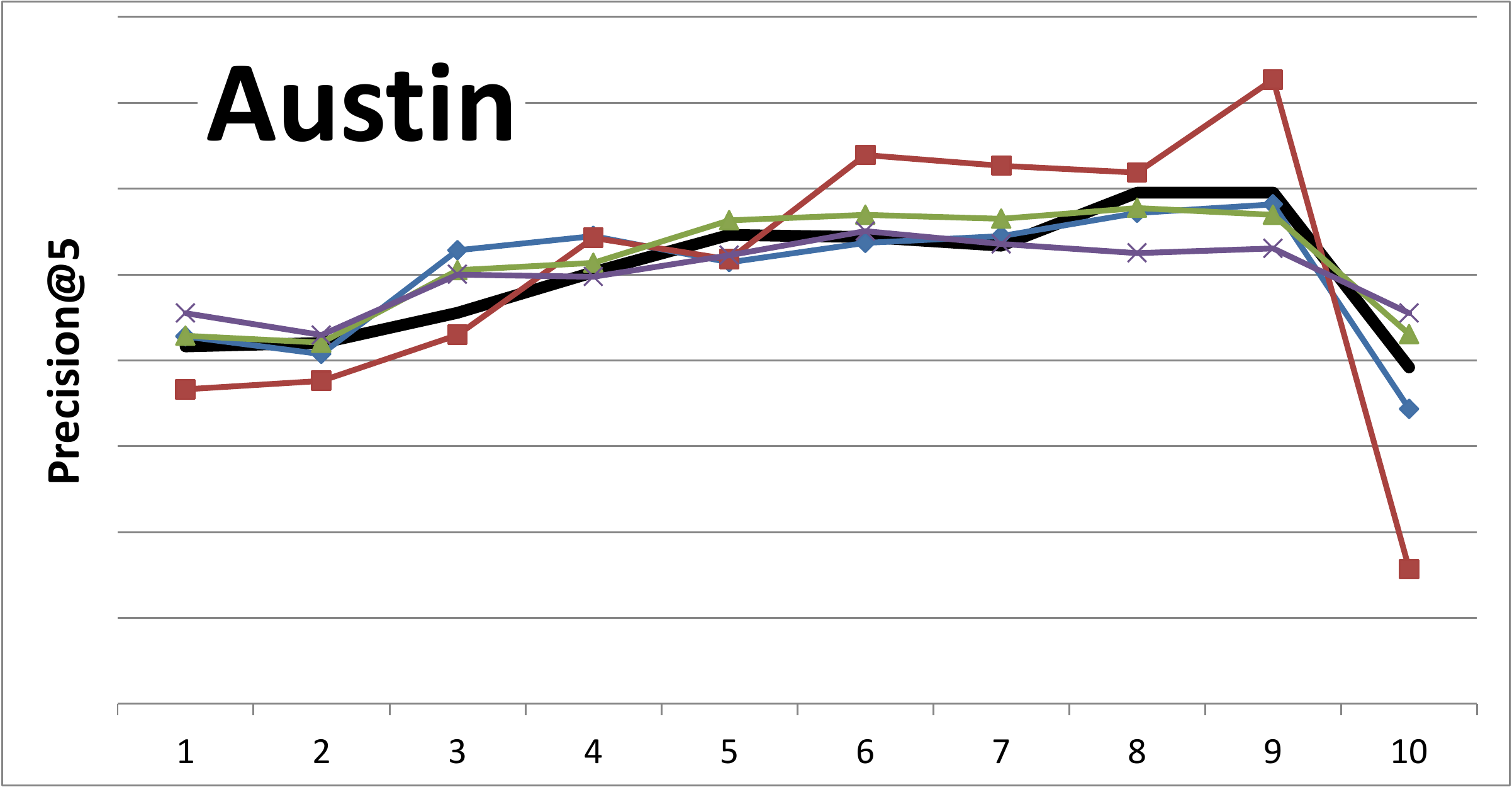}
\label{fig:datastability}
}
\caption{Stability of Results. For (a), we randomly divided the Austin data into four sets of users, each containing around 1200 users, and computed accuracy per decile removed for each set. Also shown is the same plot for all users, in bold. Since we were only interested in relative values, we centered by vertically translating each plot. For (b), we randomly divided each Austin user's data into 5 equal pieces and repeated our differential data experiment with the user hardship attribute 5 times, with the test set consisting of one of the pieces from each user and the training set consisting of the remaining four pieces. The original training and test set is in bold.}
\end{figure*}

\subsection{Stability}
In this section we examine the stability of our findings, i.e. whether the relative ranking of deciles will change with new data. We used the Austin data (the city with the most data) and the Density user hardship measure. We performed two experiments, one indicating stability with respect to different test and training sets and one indicating stability with respect to different sets of users. We found that using different sets of users was slightly less stable than using different data sets (from the same set of users), but in either case there was consistency in the less important and more important deciles. We did not study stability over time, although that would be another interesting dimension to study.

In our first experiment, we divided the users into four disjoint sets (each containing around 1200 users) and plotted accuracy versus decile removed for each set of users (see Figure~\ref{fig:userstability}). We show the plots together with a constant shift for each plot so that they are all approximately centered. We also show the plot with all users for comparison. With any of the user sets, we get a similar qualitative ranking of the deciles: the last and the closer deciles are important, and the middle deciles through Decile 9 are less important. 

In our second experiment, we examined stability with different sets of data by dividing the Austin data into 5 equal pieces and repeating our experiment 5 times with each piece serving as part of the test set for one experiment.  The resulting plots of accuracy versus decile removed all have similar shapes, but appear to be shifted relative to one another (due to the differing test sets). We show the plots in Figure~\ref{fig:datastability}, again with a constant shift. We note that for all trials, the four most important deciles were always deciles 1 through 3 and 10. The five least important deciles were always contained in deciles 4 through 9.

\subsection{Movie Ratings Dataset}
\label{sec:ML}
We consider movie recommendation as another case study. We used the well-studied MovieLens 1M dataset~\cite{MovieLens} which contains 1,000,209 anonymous ratings of 3,952 movies made by 6,040 users. Ratings in MovieLens range from one star to five stars. We divided the dataset into a training set and test set by randomly putting 20\% of the ratings for each user in a held-out test set. The remainder is the training set.

To measure accuracy, we used the common Root Mean Squared Error (RMSE) and Mean Average Error (MAE) metrics~\cite{Herlocker}, which compare the actual values in the test set with the values predicted by our algorithms. For clarity and space, we present only RMSE results; MAE results were very similar. We present results with Biased Stochastic Gradient Descent (Biased SGD), a common collaborative filtering algorithm to predict user ratings for movies in our dataset. We also used another common collaborative filtering algorithm, Alternating Least Squares (ALS), but results were again very similar so we do not present them.
The algorithms are based on a latent factor model found through matrix factorization (see~\cite{Koren} for a detailed description). We used 20 factors, so users and movies are each summarized by factor vectors of length 20. We did not search for optimal algorithms or do extensive parameter selection. Again, our goal was not to reduce error but to see which parts of the data had the most effect on error. For our algorithms, we used the implementations from~\cite{graphlab}.

\ignore{
\begin{figure}[h]
\centering
\includegraphics[width=\columnwidth]{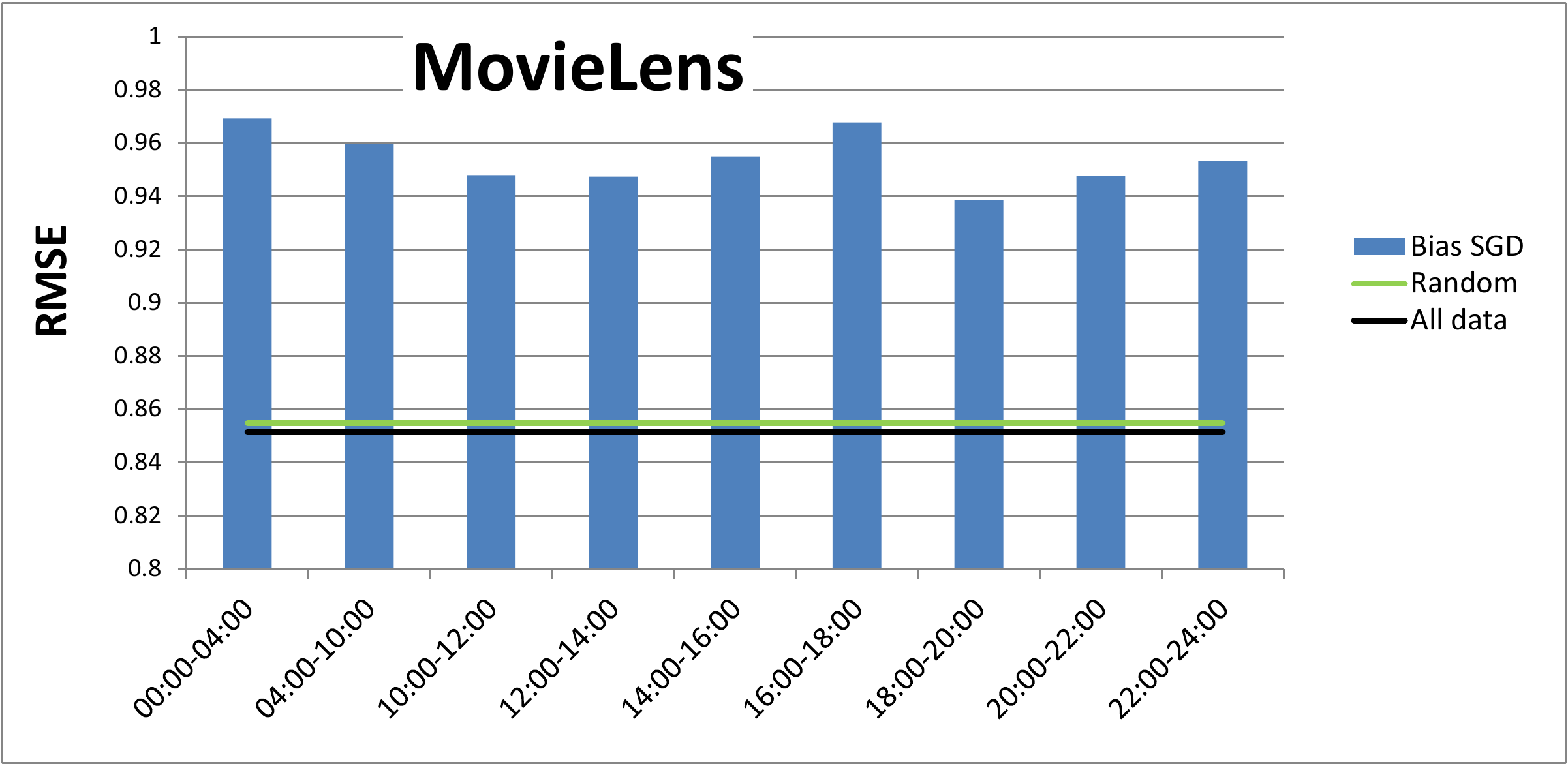}
\caption{RMSE when distinct time intervals are removed from MovieLens data. Each interval corresponds to about 10-15\% of all data. Higher RMSE values correspond to worse accuracy.}
\label{fig:MLtime}
\end{figure}

\subsubsection{Timestamp.}
The MovieLens dataset includes the timestamp of the rating (number of seconds since the epoch) and also the user's zip code, so we experimented with the effect of time of the rating on recommendation accuracy. We computed the local time of each rating and, as in Section~\ref{sec:GowallaTime}, divided the day into time periods, each with approximately 10-15\% of all the ratings, and applied our differential data analysis technique.
In this case, our technique implies the local time of the rating does not seem to have much effect on the recommendation accuracy. What might be surprising is the amount the accuracy goes down, much more than a random removal of 10-15\% of the data. The explanation is that a significant number of users almost exclusively rate movies in any given time interval, which means the recommender knows almost nothing about these users if this time period is filtered.
}

\begin{figure}[h]
\centering
\includegraphics[width=\linewidth]{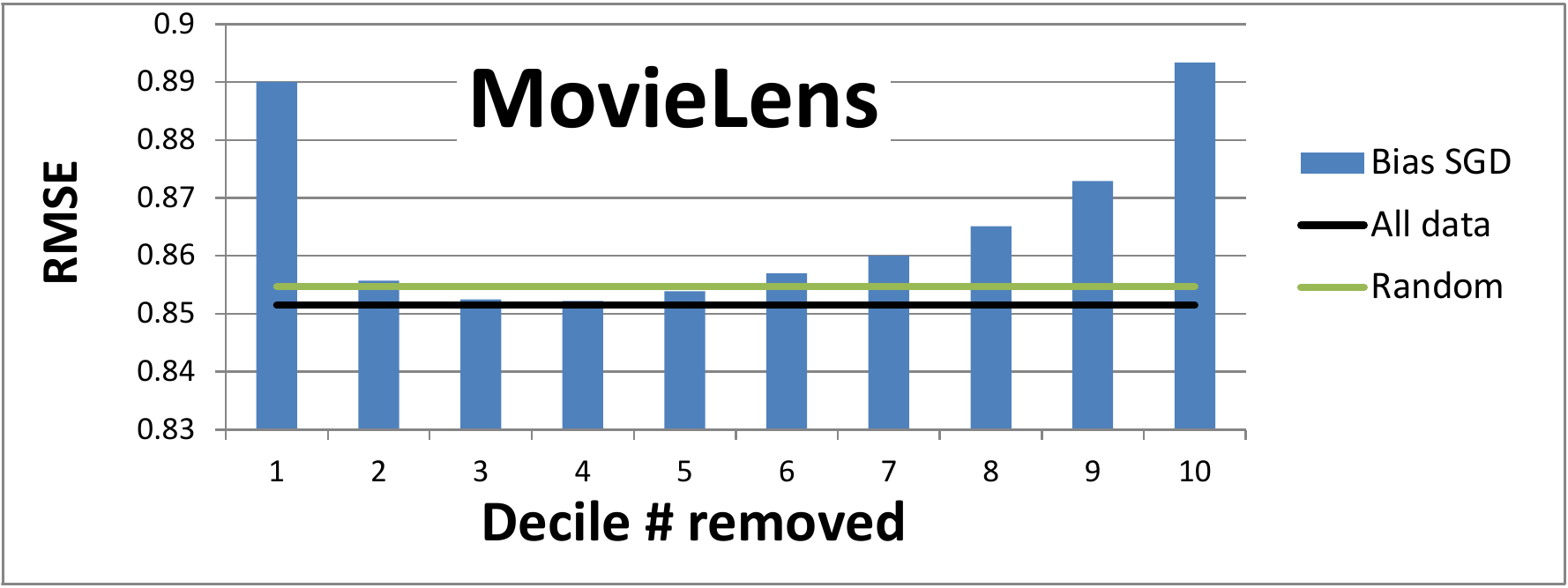}
\caption{MovieLens Data. Decile 1 contains the user's lowest ratings and Decile 10 contains the user's highest ratings. If necessary, ratings were split randomly across deciles. For example, for a particular user, both decile 1 and decile 2 may contain 1-star ratings; whether a 1-star rating goes in decile 1 or 2 is random. We also show the results with all data (no deciles deleted) and random removal of 10\% of the data.}
\label{fig:MLratings}
\end{figure}

Berkovsky et al.~\cite{Berkovsky} showed previously that high/low ratings were most important to the recommender. We validated their results using our technique of differential data analysis. We divided each user's ratings into 10 equal deciles according to value, so that Decile 1 contains the user's lowest ratings and Decile 10 contains the users high ratings. When necessary, movies with the same rating were split among the deciles in a random way. For example, if a user has rated only ten movies and given all the movies 3 stars, which movie is in Decile 1 and which is in Decile 10 is random. Figure~\ref{fig:MLratings} shows some of our results when we remove each decile in turn. We ordered each user's ratings in the training set and examined the effect of removing successive rating deciles. For example, when we removed the 10\% lowest ratings from each user, we obtained an RMSE of about $0.89$. From the figure, it is clear that the high and low ratings are the most important for recommendation accuracy. We also observe that removing deciles 3 or 4 affect recommendation accuracy the least.

\begin{figure}[h]
\centering
\includegraphics[width=\linewidth]{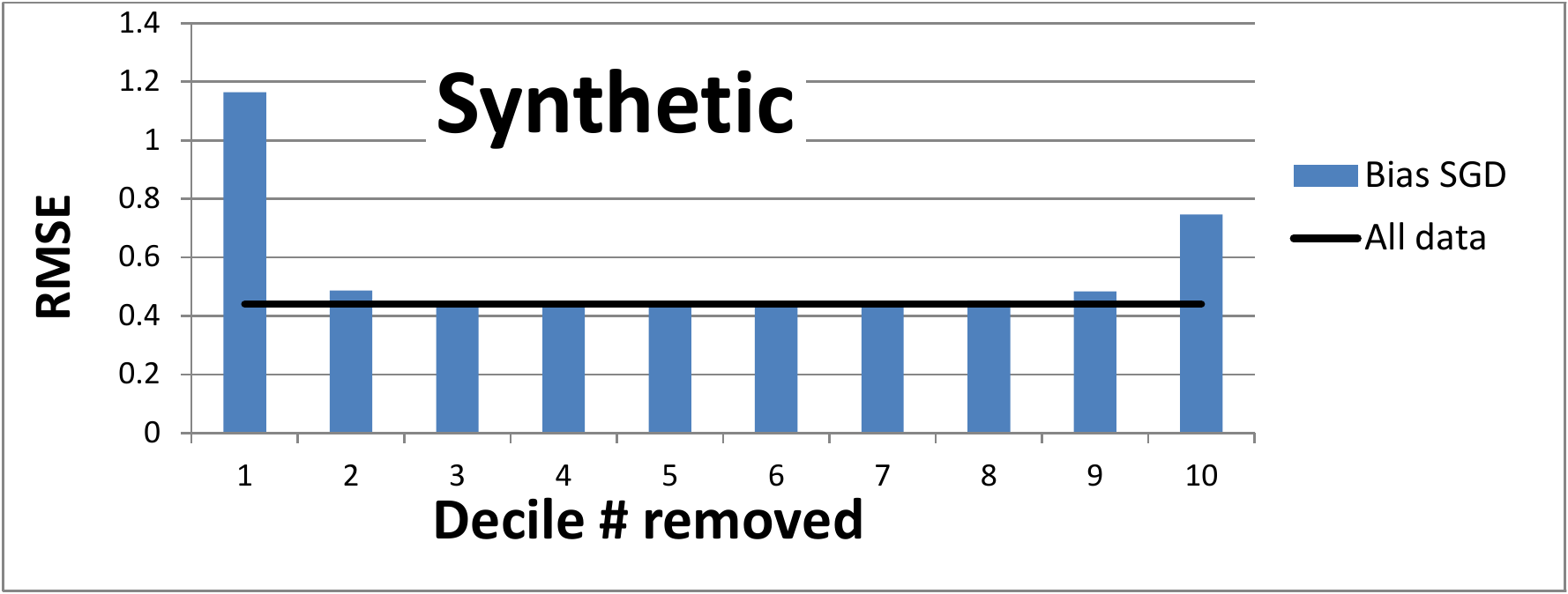}
\caption{Synthetic Data with same number of users and movies as MovieLens. When we assume a latent factor structure for the users and movies and use differential data analysis, we also see the relative importance of high/low ratings.}
\label{fig:synthratings}
\end{figure}

\subsection{Synthetic Ratings Data}
\label{sec:synthetic}
Here we investigate the generality of Berkovsky et al.'s observation. Are high and low user ratings always comparatively most important for recommendation accuracy? We generated a synthetic rating dataset, assumed to have the same number of users and movies as the MovieLens dataset. We also assumed 20 latent user factors and 20 latent movie factors. We generated random $U=6040$-by-$20$ and $V=3952$-by-$20$ matrices by selecting entries independently from a normal distribution with mean 0 and variance 1. We multiplied $UV^T$ to get the full synthetic rating matrix and randomly selected 1,000,209 ratings to get a MovieLens-like sparse dataset. We applied our technique of differential data analysis: we selected 20\% of the ratings at random as the test set and calculated recommendation accuracy after removal of each rating decile. Our results are shown in Figure~\ref{fig:synthratings}. The plot suggests that high and low ratings are indeed most important for the recommender (assuming the validity of latent factor models for the data) since this result holds even for randomly generated factor matrices and is thus a mathematical consequence of a low-dimensional ratings matrix.

\section{Applications}
We discuss here how our techniques might be applied in a recommender system in two areas: (1) user privacy through obfuscation and (2) data reduction. We show that data obfuscation or removal can perform significantly better using
our results and attributes such as user hardship (location), timestamp (location), and/or rating (movies).

\subsection{Privacy through Obfuscation}
\label{sec:obfuscation}

For ease of exposition, we discuss privacy of locations, but our techniques also apply to general item recommendations. We take a generic approach to obfuscation, instead of targeting the most sensitive locations only, because we assume the system does not know the privacy-sensitivity for a given user of any particular location. There is some experimental evidence for {\it general} trends, such as locations with lower diversity of visitors might be more sensitive and that locations visited during work hours might be less sensitive~\cite{Benisch}. Nevertheless, the privacy-sensitivity of locations is highly individual. It is an interesting question whether one can elicit a user's privacy-sensitive locations without overly burdening the user. In this work, for simplicity, we assume the system lacks any such knowledge and this is the case with most existing systems today.

Standard obfuscation strategies are adding fake ratings and suppressing ratings (for instance, see~\cite{parra-arnau}) before sending the data to the server. Fake and suppressed ratings provide plausible deniability to the user, essentially a variant on {\it randomized response}~\cite{warner}, a standard technique in surveys to gather statistics and yet protect confidentiality. For instance, for each unvisited location, with a certain probability, the user lies and says the location is visited. This way, the user can claim for any location that she did not visit it, because any visited location is potentially a lie. Similarly, suppressed ratings allow the user to claim that she was at a location even though the system shows no record of it.

The main point of this section is that these standard obfuscation techniques can be enhanced using our results. The idea is that the obfuscation can be done intelligently in a way that preserves the more important data when suppressing data and that does not interfere with the data signal when adding fake data. More concretely, for suppression we can measure the relative importance of a data chunk or interval with a z-score which is then used to guide the obfuscation process. Fixing a recommendation algorithm, for interval $i$, the z-score $z_i$ is derived using differential data analysis and is defined by:
$$z_i = \frac{a_i - a}{\sigma},$$
where $a_i$ is the accuracy with interval $i$ removed, $a$ is the average accuracy, and $\sigma$ is the standard deviation of the accuracies. For instance, analysis on Austin data as in Figure~\ref{fig:distance}, gives z-scores as shown in Table~\ref{tab:zscores}.

We assume these z-scores are considered public data that is made available to the users. With the z-scores, users can do a client-side computation to determine which data to filter or add. Calculating the z-scores of course requires a certain amount of potentially private data, which seems counter to the purpose of our approach. However, we note that this bootstrap data can be collected anonymously, as
we are calculating z-scores for a population as a whole. The bootstrap data could also come from the portion of users
who are less privacy-sensitive and willing to share their data with the server.
 
In the remainder of this section, we give results of experiments on the Gowalla dataset showing how client-side implementation of standard privacy-enhancing techniques such as suppression of ratings, adding fake ratings, and both techniques combined can be enhanced using our techniques. 

We experimented with standard and well-known obfuscation measures in this paper; our aim is to show how the efficiency of these standard obfuscation measures might be improved, not to design new obfuscation measures. For background, we list some objections to these obfuscation measures here and note that these objections are still valid even after incorporating our techniques. First, private data may still be derivable from obfuscated data, see~\cite{Zhang}. Second, the technique of fake ratings (like the k-anonymity privacy measure) suffers from a potential lack of diversity. If a vegetarian is embarrassed about his visits to steakhouses, then fake checkins consisting of all barbecue and hamburger restaurants will not provide the desired plausible deniability. Third, auxiliary data is not considered: it may be possible, for instance, to link a user's obfuscated data to the same user's unobfuscated data in another dataset (see~\cite{arvind}). Finally, these obfuscation measures are meant to obscure data points considered in isolation and do less well in hiding aggregate statistics across multiple data points; for instance, the general region where a user lives and works may still be discernible. Hence, these standard obfuscation measures are far from perfect, but there is a lack of practical alternatives.

\begin{table}
\tiny
\centering
\begin{tabular}{|c|c|c|c|c|c|c|c|c|c|}
\hline
Decile 1     & 2     & 3     & 4    & 5    & 6     & 7    & 8    & 9    & 10   \\
\hline
-1.12 & -1.07 & -0.60 & 0.04 & 0.62 &  0.58 & 0.45 & 1.27 & 1.27 & -1.45 \\
\hline
\end{tabular}
\caption{Austin z-scores for user hardship, Density measure.}
\label{tab:zscores}
\end{table}

\ignore{
\begin{table}
\scriptsize
\centering
\begin{tabular}{|c|c|}
\hline
Decile \# & z-score \\
       \hline\hline
1   & -1.12 \\
\hline
2   & -1.07 \\
\hline
3 & -0.60 \\
\hline
4 & 0.04 \\
\hline
5 & 0.62 \\
\hline
6 & 0.58 \\
\hline
7 & 0.45 \\
\hline
8 & 1.27 \\
\hline
9 & 1.27 \\
\hline
10 &-1.45 \\
\hline
\end{tabular}
\caption{Austin z-scores for user hardship, Density measure.}
\label{tab:zscores}
\end{table}
}

\begin{figure}[h]
\centering
\includegraphics[width=\columnwidth]{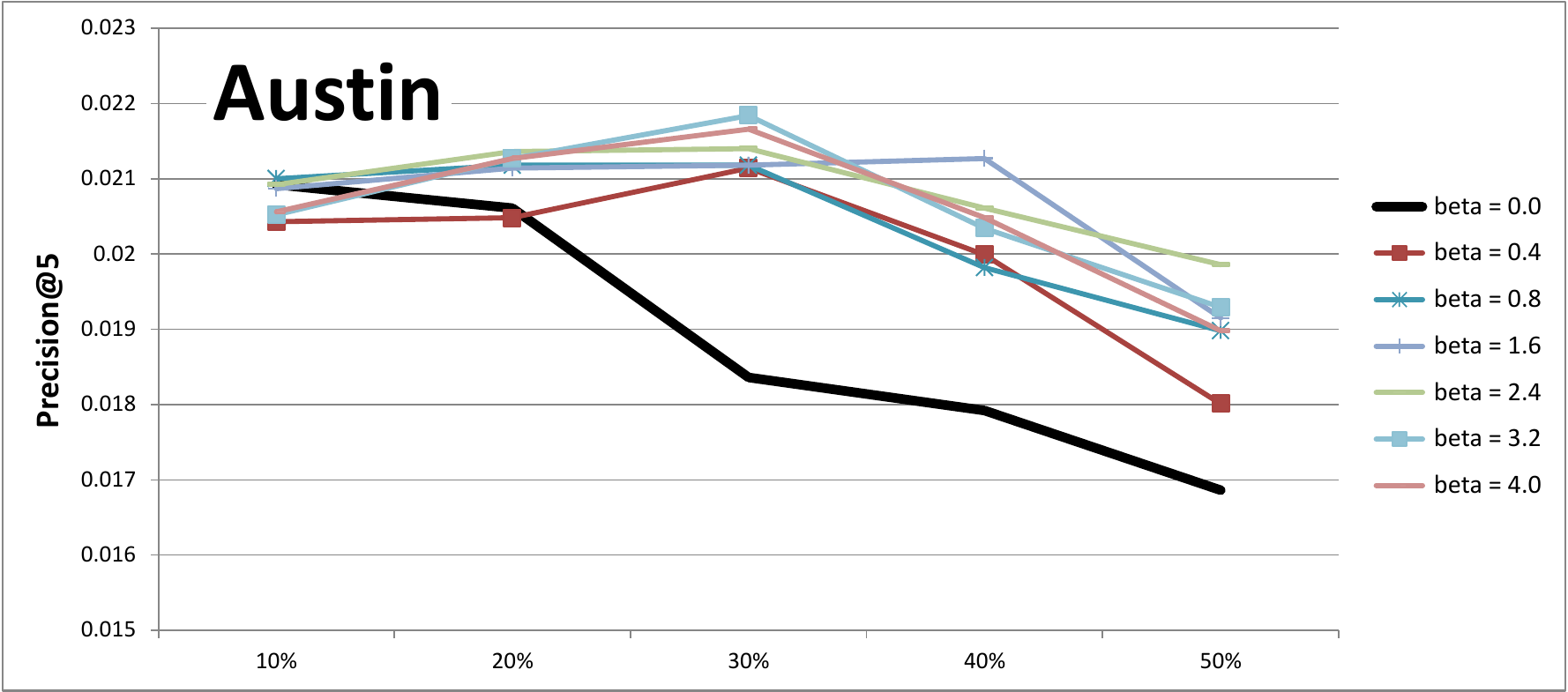}
\caption{Accuracy per percentage of data suppressed. $\beta$ is a tuning factor that weights the importance of high and low z-scores in our algorithm. The bold plot corresponds to $\beta=0.0$, or even suppression, where there is a uniform probability of any point being suppressed. For higher percentages of suppressed data, even suppression does not do as well as uneven suppression, where the probability of suppression varies depending on the decile. Note that precision@5 for no deletion is 0.0216, comparable to 30\% deletion using intelligent data suppression.}
\label{fig:smoothsuppression}
\end{figure}

\begin{figure*}[ht]
\begin{minipage}{0.24\textwidth}
\centering
\includegraphics[width=\textwidth]{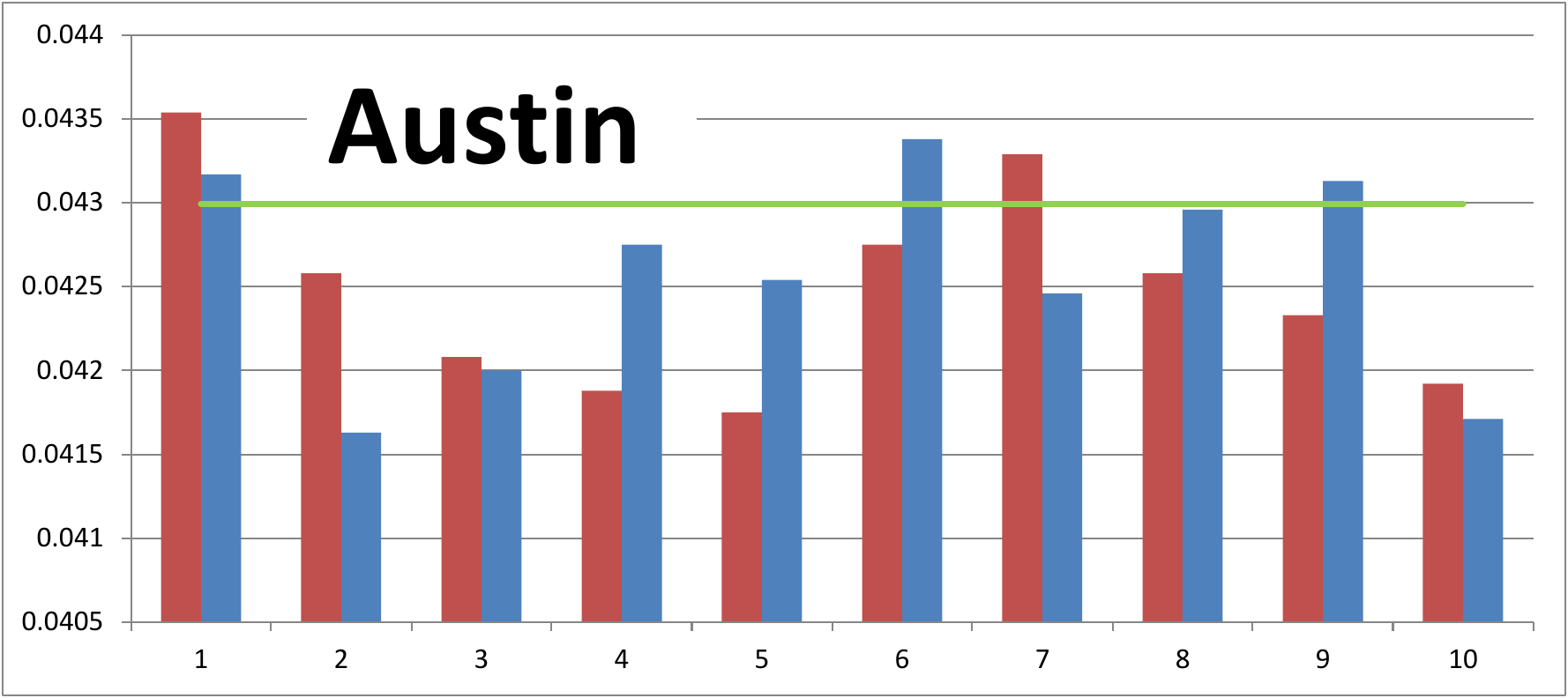}
\end{minipage}
\begin{minipage}{0.24\textwidth}
\centering
\includegraphics[width=\textwidth]{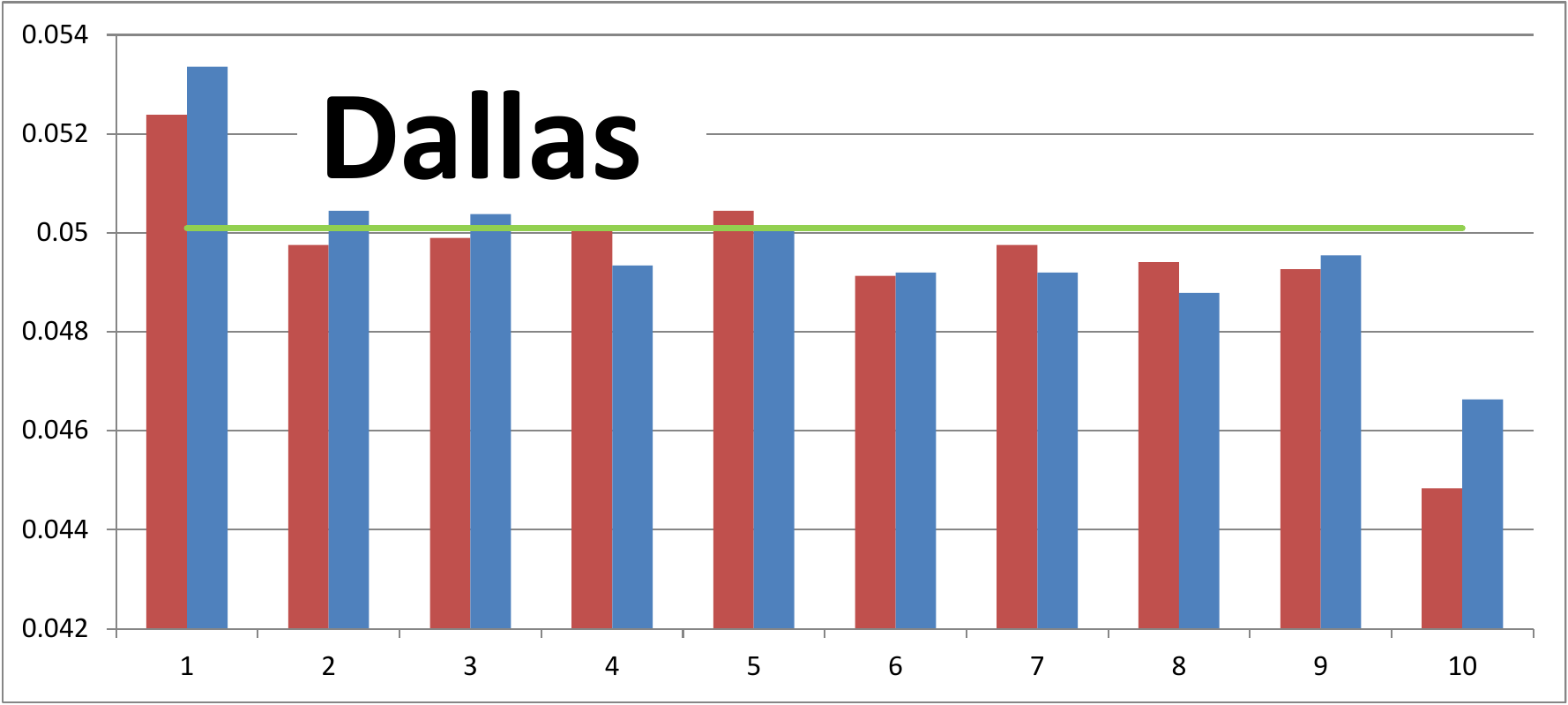}
\end{minipage}
\begin{minipage}{0.24\textwidth}
\centering
\includegraphics[width=\textwidth]{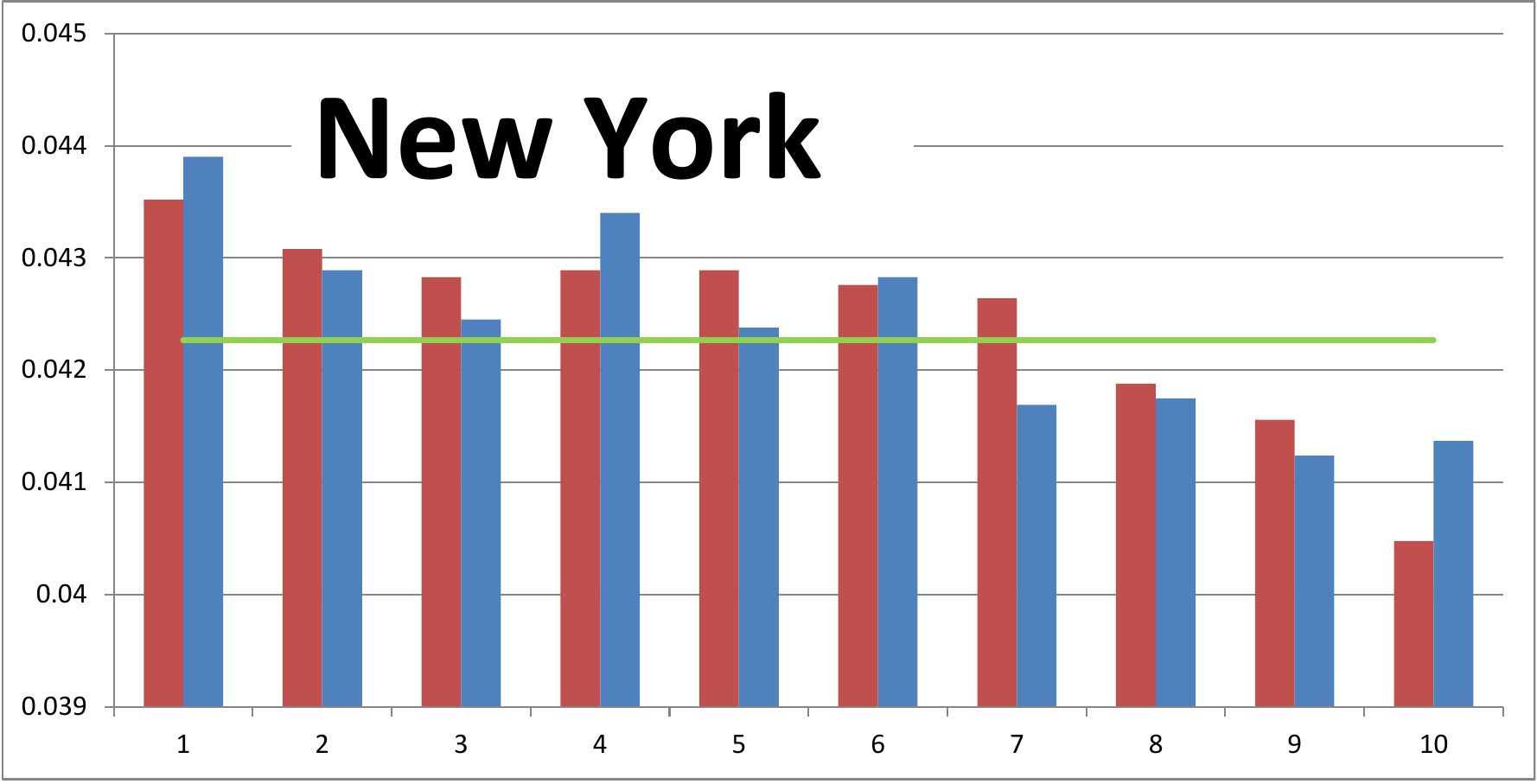}
\end{minipage}
\begin{minipage}{0.24\textwidth}
\centering
\includegraphics[width=\textwidth]{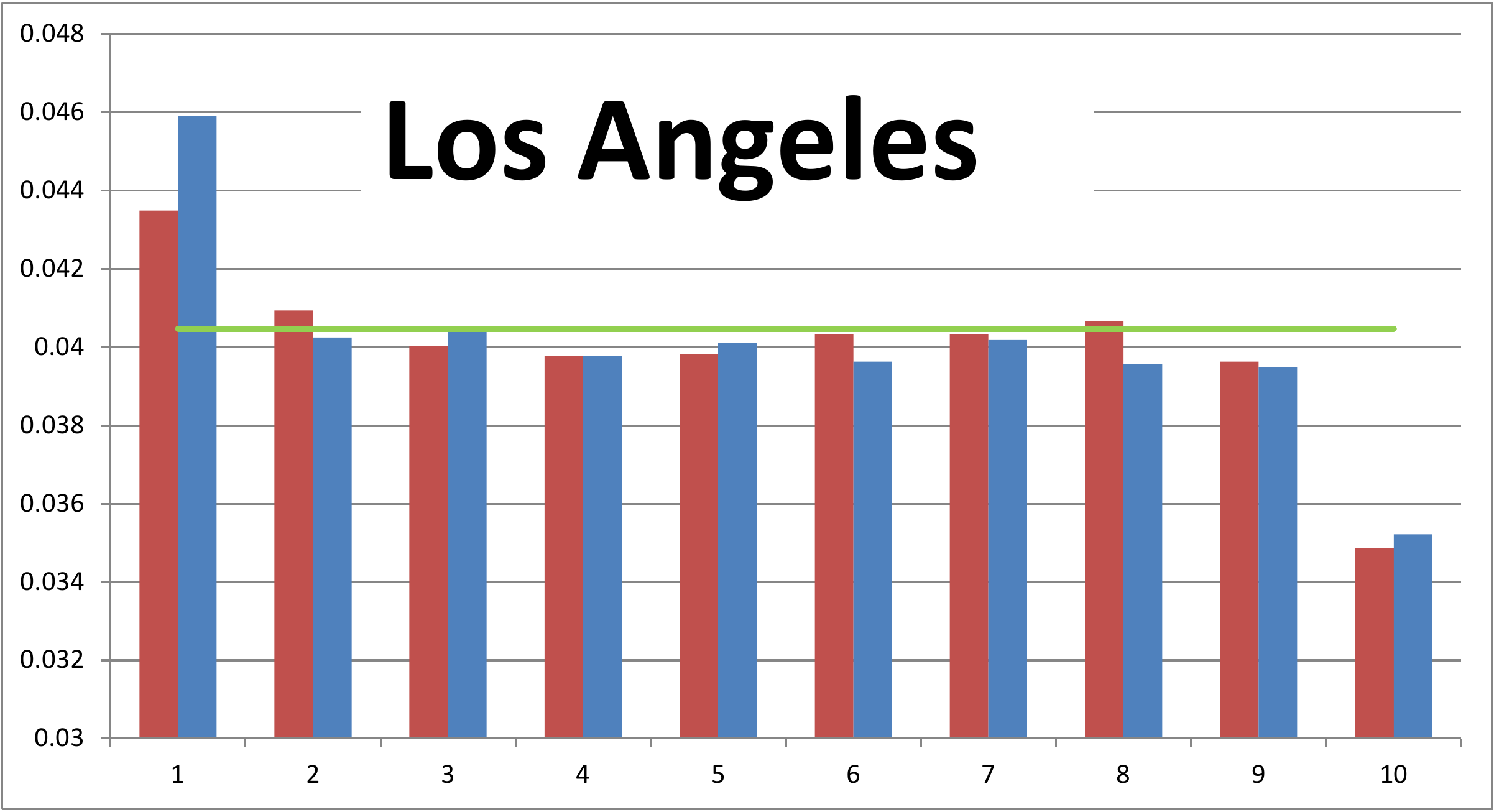}
\end{minipage}

\begin{minipage}{\textwidth}
\centering
\includegraphics[width=0.20\textwidth]{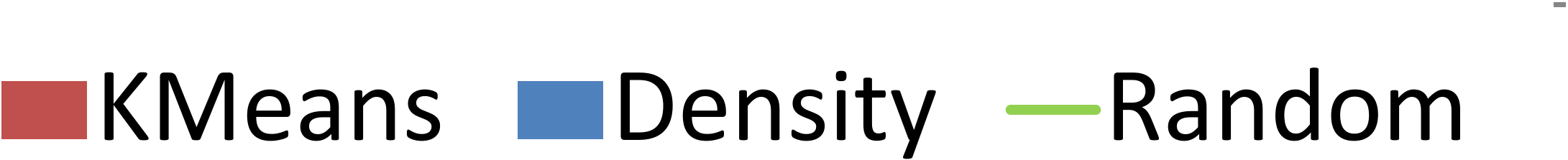}
\end{minipage}
\caption{Recommendation accuracy when adding 100\% fake data as a function of decile. We generated fake ratings for each user in an amount ten times his actual ratings. Ranking these ratings and applying differential data analysis shows that recommendation accuracy is affected least by the fake data with least user hardship values.}
\label{fig:Fake}
\end{figure*}

{\noindent \bf Intelligent Data Suppression.} Here we discuss intelligent data suppression using our techniques.
In Figure~\ref{fig:smoothsuppression} we show the results of a simple algorithm for optimizing which data to suppress. We choose an overall suppression level $\alpha$ and a system parameter $\beta$ which tunes the importance of high and low z-scores. Then, for a decile with z-score $z_s$, let $t = e^{\beta z_s}/(e^{\beta z_s} + e^{-\beta z_s})$. We suppress ratings according to the following formula:
$$
\hat{p}(z_s) = \left\{
\begin{array}{l l}
2\alpha t  & \quad \textrm{if $t < 1/2$}\\
2(1-\alpha)t + 2\alpha -1 & \quad \textrm{if $t \geq 1/2$}
\end{array} \right.
$$
The formula was designed so that $\hat{p} \rightarrow 1$ as $z_s$ becomes large (positive), $\hat{p} \rightarrow 0$ as $z_s$ becomes large (negative), and if $z_s$ is 0, then $\hat{p}$ has the value $\alpha$, the overall suppression level. Finally, we normalize by $p=\hat{p}/k$, where the constant k is chosen so that the average suppression level over all deciles is $\alpha$, i.e., $\frac{1}{10} \sum \hat{p} = \alpha$. For each data point in the decile, we suppress, or delete, the point with probability $p$. We call this uneven obfuscation, as the probability of deletion depends on the decile a point belongs to; deciles considered less important to the recommender have a higher suppression probability.

For the experiment in Figure~\ref{fig:smoothsuppression}, we divided the Austin data into two halves, Half A and Half B. Half A can be thought of as training data and Half B as test data. Half A itself was divided into 80\% training and 20\% test in order to apply differential data analysis to calculate z-scores, which were then used in the suppression experiment on Half B. Figure~\ref{fig:smoothsuppression} shows uneven obfuscation does markedly better than even obfuscation for overall suppression levels larger than 10\%. Values of $\beta$ around 3 or larger do best. Large values of $\beta$ correspond to a strategy of simply deleting only from the intervals with positive z-scores.

{\noindent \bf Fake Data.} In addition to suppressing actual data, our differential data analysis techniques can also be used for effectively adding fake data.  
In Figure~\ref{fig:Fake}, we took the Gowalla training data and created random fake data, generating for each user 10 times the number of his actual checkins. We applied our differential data analysis technique to determine which fake data impacts recommendation accuracy most as follows. We divided each user's fake data into deciles using our two user hardship measures, the Density measure (distance to actual data) and the KMeans measure (distance to KMeans centroids). We added each decile in turn to see the effect on recommender accuracy (using our held out test data). Each decile of a user's fake data contains the same number of data points as all his actual data. Figure~\ref{fig:Fake} shows the results for Austin, New York, Dallas, and LA. Dallas, LA, and New York are consistent. Austin is a little different (as with data suppression, see Figure~\ref{fig:distance}), but still similar.

The results say that fake data with small user hardship values interferes least with recommendation accuracy. This may be surprising at first, since nearby businesses may be quite diverse. For instance, fake data near an actual checkin to a Mexican restaurant may consist of a bike shop, a drug store, and a flower shop. One intuitive explanation for our result is that there is a tendency for businesses near each other to target like-minded consumers, for instance the ``Whole Foods effect,''~\cite{wholefoods} in which an entire shopping center attracts affluent, health-oriented consumers.

Note that with data suppression of amounts greater than 10\% we must spread the suppression across multiple deciles, hence the intelligent data suppression algorithms in the previous section. In contrast, there is no need to take fake data from multiple deciles, since in practice as much fake data as desired can be generated in any decile. However, if fake data is all chosen from the same decile, it is possible that using geometric arguments one can infer some aspects of the real data. We did not experiment in this direction, but if
this is a concern one can trade some privacy for accuracy by mixing the deciles and implementing intelligent data faking algorithms, e.g. half of the fake data from the 9th decile and half from the other deciles.

\begin{figure*}[ht]
\centering
\subfloat[][User hardship]
{\includegraphics[width=0.30\textwidth]{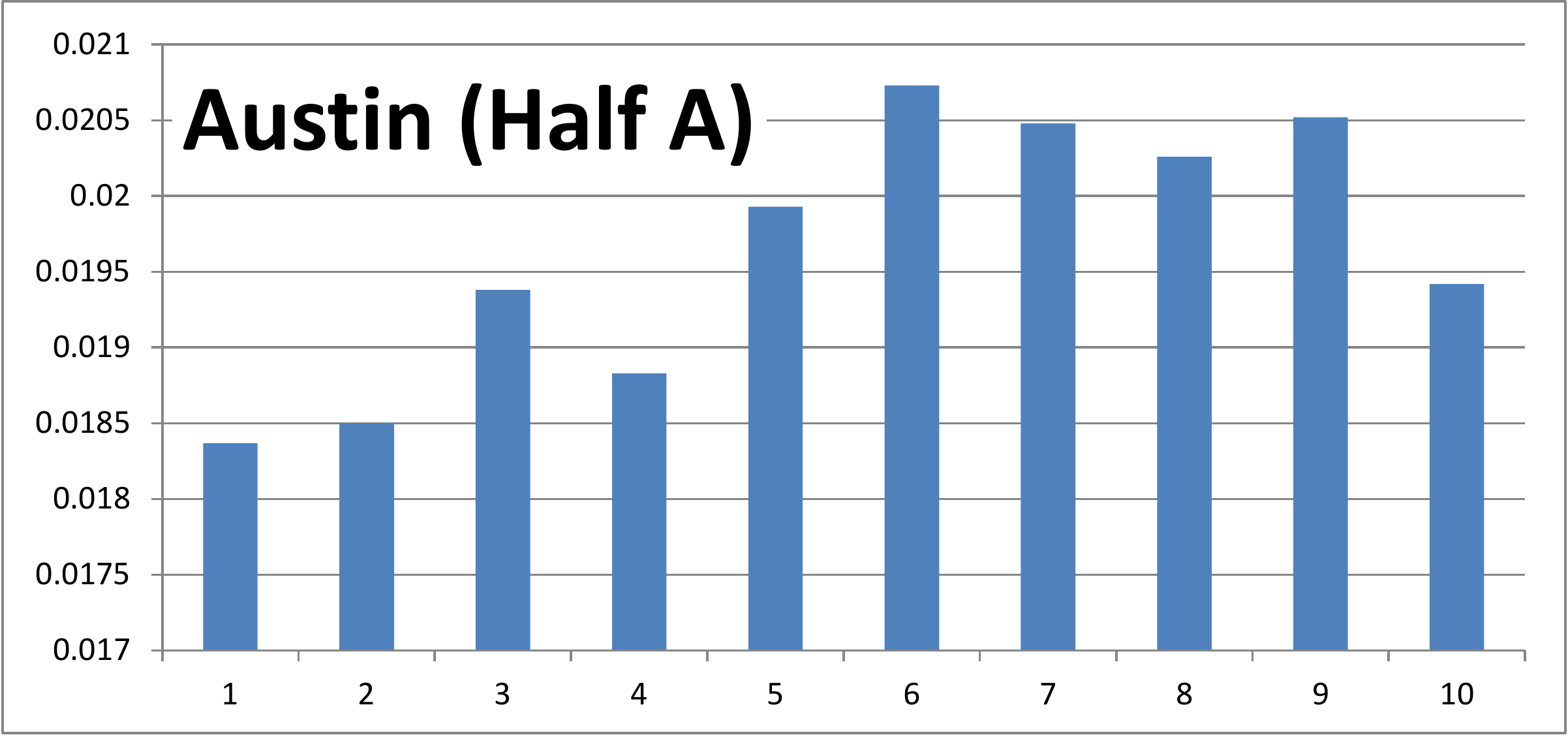}
\label{fig:halfASuppress}}
\qquad
\subfloat[][50\% fake data]
{
\includegraphics[width=0.30\textwidth]{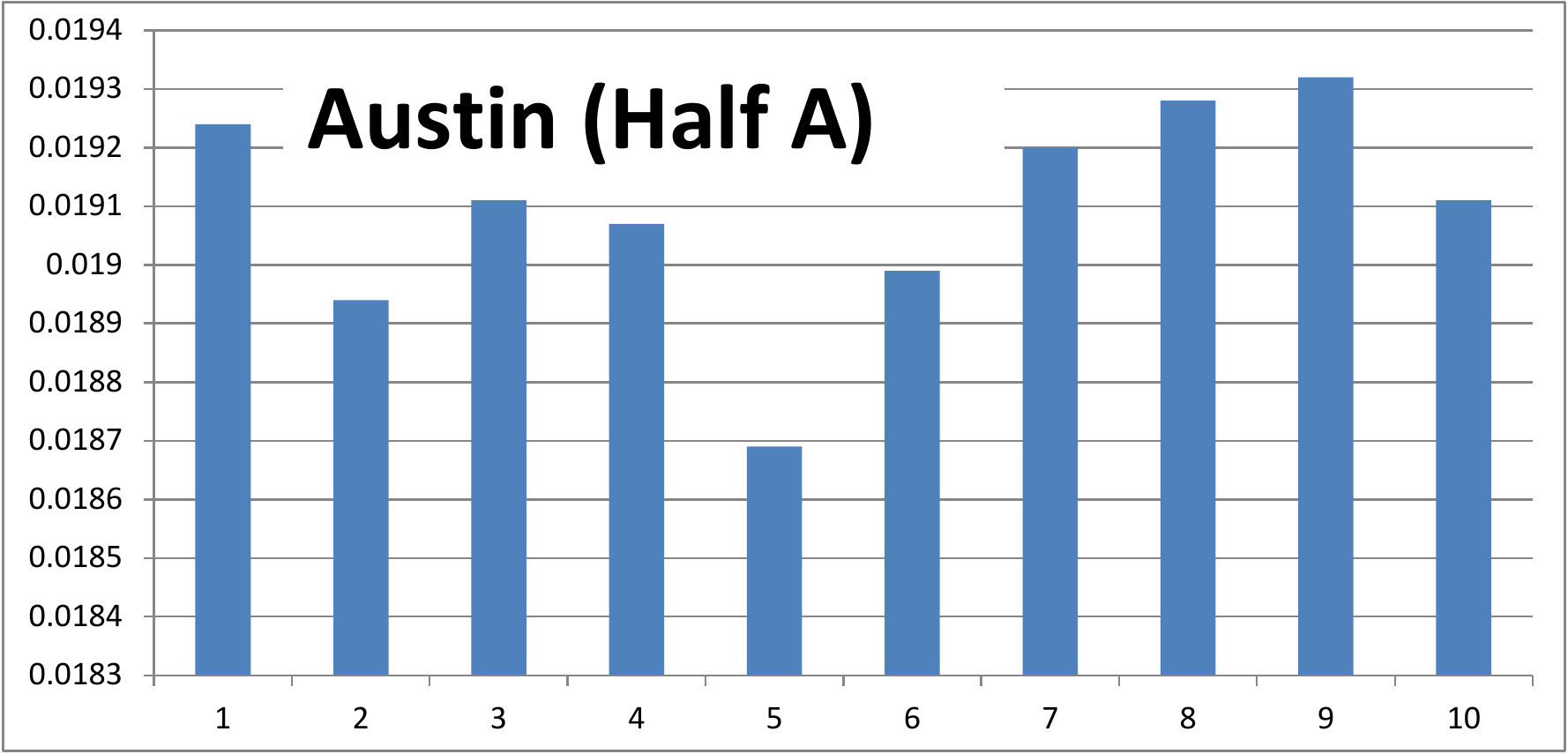}
\label{fig:halfAFake}
}
\qquad
\subfloat[][Timestamp]
{
\includegraphics[width=0.25\textwidth]{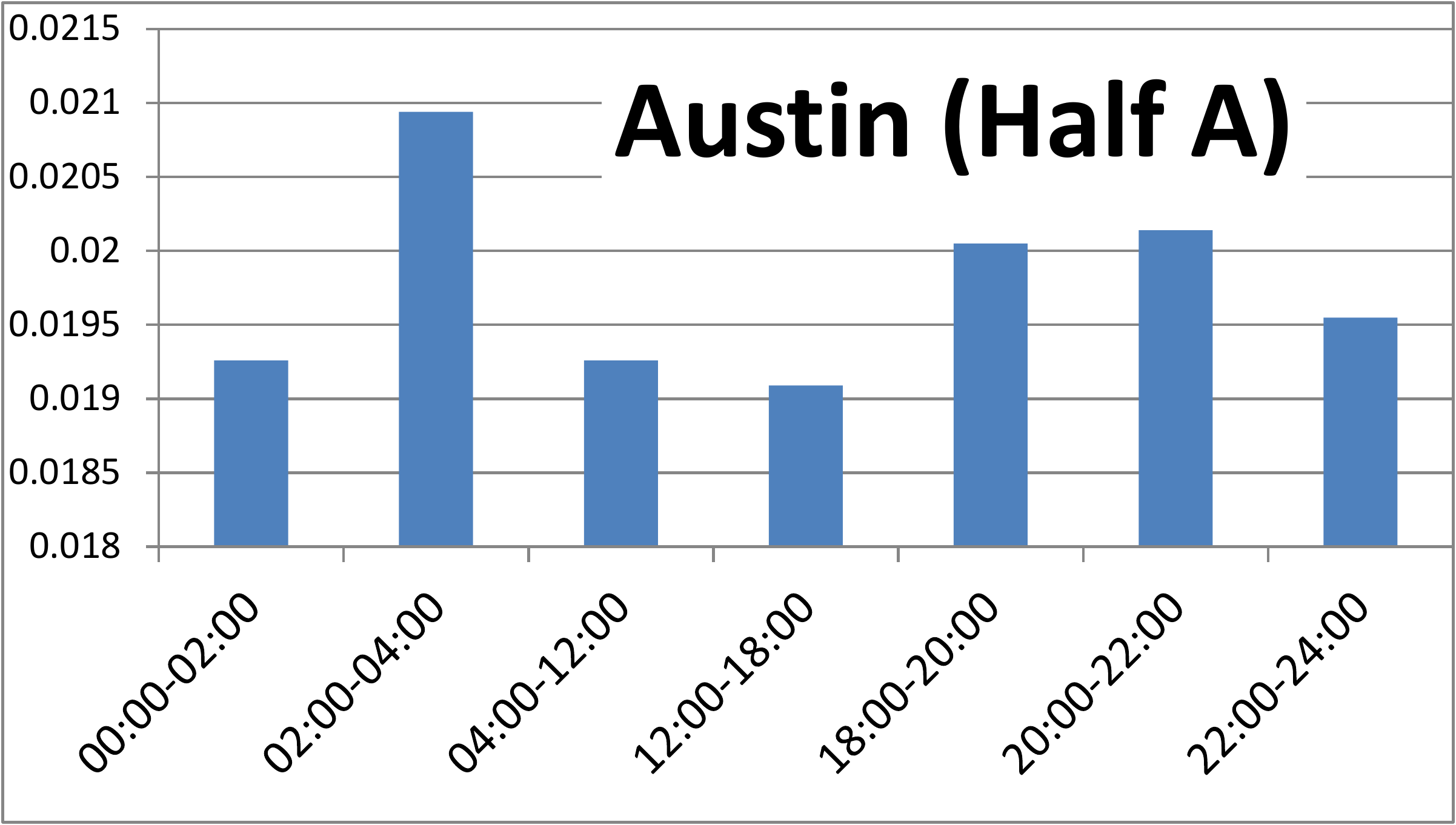}
\label{fig:halfATime}
}
\caption{Results of differential analysis on Austin Half A. These results were applied to Austin Half B in order to efficiently remove data and replace with fake, achieving significantly higher recommendation accuracy compared to random suppression and replacement. These results were also applied to demonstrate data reduction on Austin Half B.}
\end{figure*}

{\noindent \bf Combined data suppression and faking.} Finally, we experimented with a more realistic obfuscation scenario, combining data suppression and fake data. We considered the case
of a replacement strategy where each actual rating removed is replaced with one fake rating. We divided the Austin Gowalla data into a training half and a test half, Half A and Half B.

To remove ratings, we used differential data analysis on Half A to generate z-scores (see Figure~\ref{fig:halfASuppress}) for each user hardship decile (Density measure).  We used these z-scores on Half B (with $\beta=3.0$) to reduce data from percentages ranging from 10\% to 50\%.

\begin{figure}[htbp]
\centering
\includegraphics[width=\columnwidth]{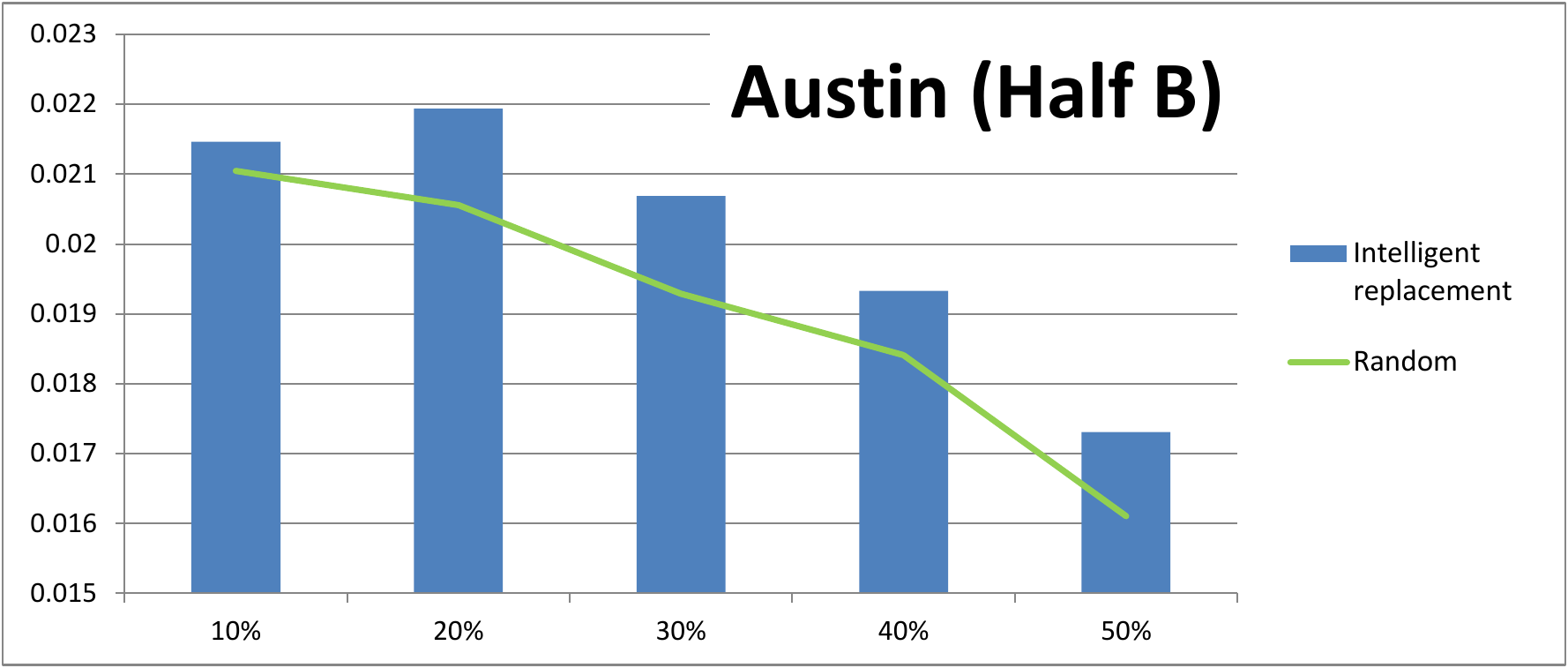}
\caption{Intelligent replacement for Half A of the Gowalla dataset, where removal and substitution is done intelligently according to differential analysis on Half B. We compare accuracy to random replacement.}
\label{fig:replacement}
\end{figure}

To add fake ratings in the same amounts to Half B, we generated five times the amount of original data for Half A and divided into deciles according to user hardship (Density measure). We obtained a plot as in Figure~\ref{fig:halfAFake}. This data for Half A suggested that fake data in the 9th decile affects accuracy the least, so we generated random data for Half B (one times to five times the original amount) and added the 9th decile of this fake data to Half B. In this way, we have replaced differing percentages of the original data with fake data. Our results are shown in Figure~\ref{fig:replacement}. We obtain significantly higher accuracy compared to random replacement.

\subsection{Data Reduction}
\begin{figure}[htbp]
\centering
\includegraphics[width=\columnwidth]{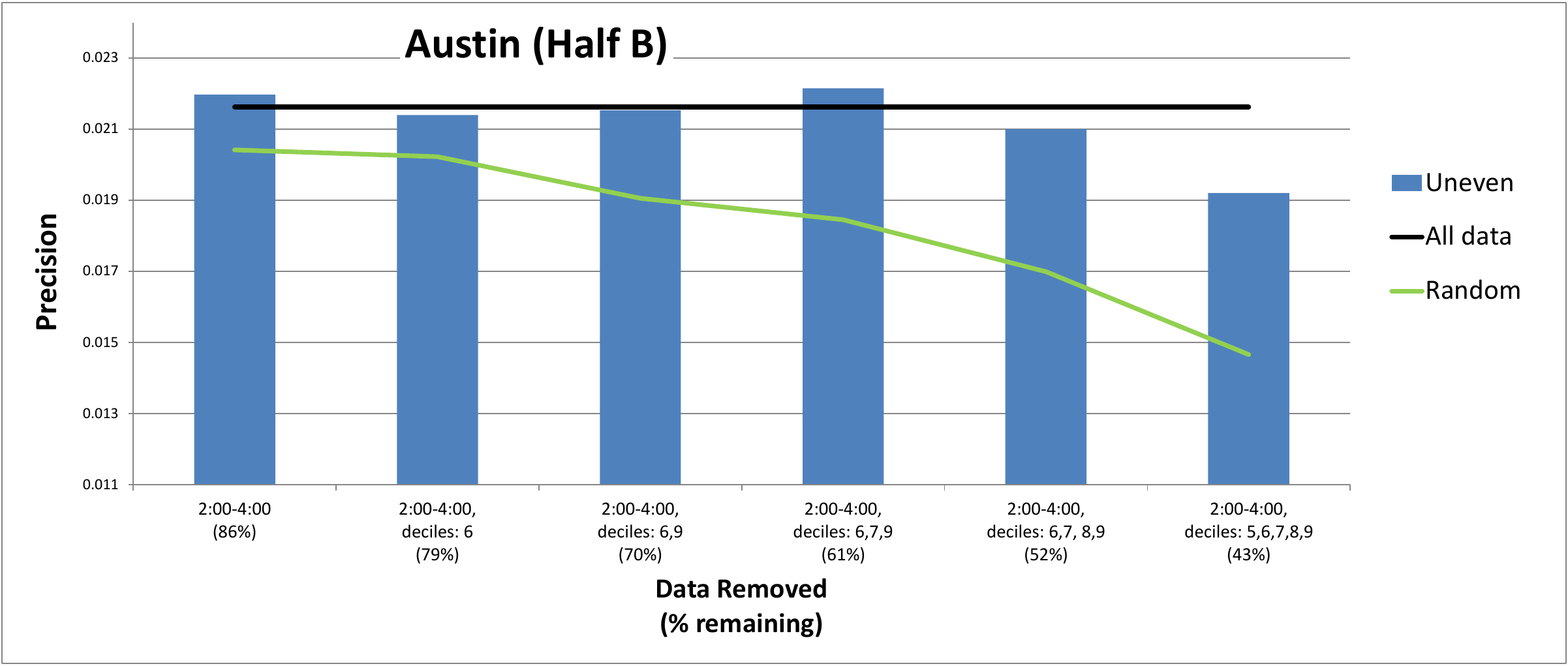}
\caption{Recommendation accuracy for the Gowalla dataset as increasing numbers of user hardship deciles are removed from the training set. The time interval from 2 am -- 4 am is removed initially. Recommendation accuracy can be maintained while removing approximately 40\% of the data.}
\label{fig:SuccessiveRemoval}
\end{figure}

Even when users are willing to share data with the server, our techniques have potential applications in data savings. Here we are suggesting that a recommender system can use our techniques to purge data that is less useful or even confusing to the system. In Figure~\ref{fig:SuccessiveRemoval}, we show results for an experiment with the Gowalla dataset for the city of Austin. Again, we divided the Austin Gowalla data into a training half and a test half, Half A and Half B. We used Half A to determine which data to remove from Half B. Consulting Figures~\ref{fig:halfASuppress} and~\ref{fig:halfATime}, we see the 2 am - 4 am time period was noise for Austin Half A, so we removed it from Half B. Then, we successively removed the user hardship deciles (Density measure) contributing the least to recommendation accuracy of Half A. We were able to remove approximately 40\% of the training data from Half B and still maintain recommendation accuracy comparable to the accuracy with all the training data.

\section{Related Work}
A natural approach to enhance privacy of recommender data is to obfuscate data before uploading to the server by adding noise, e.g.~\cite{parra-arnau, spotme}. As shown originally in~\cite{PolatDu}, collaborative filtering algorithms can still work well on obfuscated data, as the noise effects will go away with a large number of users. We view obfuscation as complementary to this work - we illustrate in Section~\ref{sec:obfuscation} how the efficiency of obfuscation can be improved with the techniques in this paper. This approach is similar to that in~\cite{Zhang_EC}, where the server sends the client a linear mapping to apply to his user ratings vector, essentially applying uneven obfuscation. Also close to the spirit of our work and also in the context of obfuscation,~\cite{Berkovsky} recognized the importance of high and low user ratings for the recommender system, noting that high and low ratings were most useful in the sense that their system was less tolerant of obfuscating the extreme ratings.

We contrast this previous work with ours in two ways. First, our techniques apply to not just user rating data. In particular, location data comes with a distance measure that is absent from, say, movie ratings data, and we use this distance measure as an integral part of our technique. Second, we give intuitive, easy-to-understand characterizations of which data is most important to the recommender system, unlike~\cite{Zhang_EC} where the characterization is abstract.

A body of work for location privacy concentrates on localizing the user at certain points of time, e.g., see~\cite{Krumm} for a survey. The data generally consists of timestamped location checkins, and the goal is to identify or link these points. For our problem of privacy-preserving location recommendation, we consider the location checkins to be sporadic (even without a timestamp), so the privacy concerns are more about each location in isolation. However, for both this body of work and ours, the distance metric on location checkins is of central importance. For this body of work, distance is an integral part of obfuscation and inference algorithms. For us, distance is a critical ingredient for evaluating relative importance of data points to the recommender system.

Differential privacy is a server-side approach to protecting individual privacy~\cite{dwork} where data is obfuscated by the server so that the presence or absence of individual records cannot be inferred. This is a topic unrelated to the subject of this paper, despite our use of the word ``differential'' and our application to privacy.

\section{Conclusion and Future Work}
We presented a simple technique called differential data analysis. Using a dataset of Gowalla checkins and a novel data attribute called user hardship,
we found that locations closer to a user's usual haunts were most important and locations further away were less important: low- to mid-user hardship is thus the optimum zone for discovering user preferences. We see the concept of user hardship as applying to other types of data besides location data. For instance, activity data can be classified according to difficulty or resources required (mountain biking vs.\ going to a movie), and it would be interesting to do a similar analysis with an activity recommender.
We also confirmed previous work with the MovieLens dataset that showed high and low user ratings are most important to the recommender.
For timestamp, using the Gowalla checkins, we found that in general very late-night data is least useful and may even confuse the recommender.
Interestingly, our results differ from city to city. The root causes for these differences deserve further study. Many other data attributes are also worth exploring. One example is social network data. Such graph-based data has become a common ingredient in recommender algorithms, and it would be interesting to explore which part of the graph is actually important to the recommender.

Our work has applications to user privacy and data reduction. One of the main challenges of recommender systems is user privacy, as the data to build a user profile can be potentially sensitive or embarrassing. Adding fake data and suppressing actual data are standard tactics to enhance user privacy, and one consequence of our work is to make these tactics more efficient. For instance, we proposed removing locations and adding fake locations selectively to maintain recommender accuracy.

\bibliographystyle{abbrv}
\scriptsize
\bibliography{refs}

\begin{thebibliography}{10}

\bibitem{agrawal}
R.~Agrawal and R.~Srikant.
\newblock Privacy-preserving data mining.
\newblock In {\em SIGMOD Conference}, pages 439--450, 2000.

\bibitem{awad}
N.~F. Awad and M.~S. Krishnan.
\newblock The personalization privacy paradox: An empirical evaluation of
  information transparency and the willingness to be profiled online for
  personalization.
\newblock {\em MIS Quarterly}, 30(1):13--28, 2006.

\bibitem{Benisch}
M.~Benisch, P.~G. Kelley, N.~Sadeh, and L.~F. Cranor.
\newblock Capturing location-privacy preferences: quantifying accuracy and
  user-burden tradeoffs.
\newblock {\em Personal Ubiquitous Comput.}, 15(7):679--694, Oct. 2011.

\bibitem{Berkovsky}
S.~Berkovsky, Y.~Eytani, T.~Kuflik, and F.~Ricci.
\newblock Enhancing privacy and preserving accuracy of a distributed
  collaborative filtering.
\newblock RecSys '07, pages 9--16, New York, NY, USA, 2007.

\bibitem{Locaccino}
{Carnegie Mellon University}.
\newblock Locaccino.
\newblock \url{http://locaccino.org/}.

\bibitem{wholefoods}
W.~Doig.
\newblock {Whole Foods is coming? Time to buy}.
\newblock
  \url{http://www.salon.com/2012/05/05/whole_foods_is_coming_time_to_buy/}.

\bibitem{dwork}
C.~Dwork.
\newblock A firm foundation for private data analysis.
\newblock {\em Commun. ACM}, 54(1):86--95, 2011.

\bibitem{BigDataStorage}
K.~Fogarty.
\newblock Big data means big storage choices.
\newblock
  \url{http://www.informationweek.com/storage/systems/big-data-means-big-storage-choices/240004436}.

\bibitem{PEW2000}
S.~Fox.
\newblock Trust and privacy online: Why americans want to rewrite the rules.
\newblock
  \url{http://www.pewinternet.org/~/media//Files/Reports/2000/PIP_Trust_Privacy_Report.pdf.pdf},
  2000.

\bibitem{Herlocker}
J.~L. Herlocker, J.~A. Konstan, L.~G. Terveen, and J.~T. Riedl.
\newblock Evaluating collaborative filtering recommender systems.
\newblock {\em ACM Trans. Inf. Syst.}, 22(1):5--53, Jan. 2004.

\bibitem{PEW2}
M.~M. Jan~Lauren, Aaron~Smith.
\newblock Privacy and data management on mobile devices.
\newblock
  \url{http://pewinternet.org/~/media//Files/Reports/2012/PIP_MobilePrivacyManagement.pdf}.

\bibitem{Koren}
Y.~Koren, R.~Bell, and C.~Volinsky.
\newblock Matrix factorization techniques for recommender systems.
\newblock {\em Computer}, 42(8):30--37, Aug. 2009.

\bibitem{Krumm}
J.~Krumm.
\newblock A survey of computational location privacy.
\newblock {\em Personal Ubiquitous Comput.}, 13(6):391--399, Aug. 2009.

\bibitem{graphlab}
Y.~Low, J.~Gonzalez, A.~Kyrola, D.~Bickson, C.~Guestrin, and J.~M. Hellerstein.
\newblock Graphlab: A new framework for parallel machine learning.
\newblock {\em CoRR}, abs/1006.4990, 2010.

\bibitem{arvind}
A.~Narayanan and V.~Shmatikov.
\newblock Robust de-anonymization of large sparse datasets.
\newblock In {\em IEEE Symposium on Security and Privacy}, pages 111--125,
  2008.

\bibitem{parra-arnau}
J.~Parra-Arnau, D.~Rebollo-Monedero, and J.~Forn{\'e}.
\newblock A privacy-protecting architecture for collaborative filtering via
  forgery and suppression of ratings.
\newblock In {\em DPM/SETOP}, pages 42--57, 2011.

\bibitem{PolatDu}
H.~Polat and W.~Du.
\newblock Privacy-preserving collaborative filtering using randomized
  perturbation techniques.
\newblock ICDM '03, pages 625--628, Washington, DC, USA, 2003. IEEE Computer
  Society.

\bibitem{spotme}
D.~Quercia, I.~Leontiadis, L.~McNamara, C.~Mascolo, and J.~Crowcroft.
\newblock {SpotME If You Can: Randomized Responses for Location Obfuscation on
  Mobile Phones}.
\newblock ICDCS, Minneapolis, USA, June 2011.

\bibitem{shardanand}
U.~Shardanand and P.~Maes.
\newblock Social information filtering: Algorithms for automating "word of
  mouth".
\newblock In {\em CHI}, pages 210--217, 1995.

\bibitem{Toch}
E.~Toch, J.~Cranshaw, P.~H. Drielsma, J.~Y. Tsai, P.~G. Kelley, J.~Springfield,
  L.~Cranor, J.~Hong, and N.~Sadeh.
\newblock Empirical models of privacy in location sharing.
\newblock Ubicomp '10, pages 129--138, New York, NY, USA, 2010. ACM.

\bibitem{MovieLens}
{University of Minnesota}.
\newblock Movielens.
\newblock \url{http://movielens.umn.edu/}.

\bibitem{warner}
S.~L. Warner.
\newblock Randomized response: A survey technique for eliminating evasive
  answer bias.
\newblock In {\em Journal of the American Statistical Association}, volume
  60(309), pages 63--69, 1965.

\bibitem{Ye}
M.~Ye, P.~Yin, W.-C. Lee, and D.-L. Lee.
\newblock Exploiting geographical influence for collaborative point-of-interest
  recommendation.
\newblock SIGIR '11, pages 325--334, New York, NY, USA, 2011. ACM.

\bibitem{Zhang}
S.~Zhang, J.~Ford, and F.~Makedon.
\newblock Deriving private information from randomly perturbed ratings.
\newblock In {\em Siam Conference on Data Mining}, 2006.

\bibitem{Zhang_EC}
S.~Zhang, J.~Ford, and F.~Makedon.
\newblock A privacy-preserving collaborative filtering scheme with two-way
  communication.
\newblock In {\em EC}, pages 316--323, 2006.

\end{thebibliography}
\end{document}